\documentclass[preprint]{aastex62}
\received{October , 2018}
\revised{November , 2018}
\accepted{}
\begin{document}

\author{Jean-Pierre~Roques}
\affil{ CNRS; IRAP; 9 Av. colonel Roche, BP 44346, F-31028 Toulouse cedex 4, France\\
 Universit\'e de Toulouse; UPS-OMP; IRAP;  Toulouse, France\\}
\author{Elisabeth Jourdain}
\affil{ CNRS; IRAP; 9 Av. colonel Roche, BP 44346, F-31028 Toulouse cedex 4, France\\
 Universit\'e de Toulouse; UPS-OMP; IRAP;  Toulouse, France\\}
\title{On the high-energy emissions of compact objects observed with \textit{INTEGRAL} SPI:  Event selection impact on source spectra and scientific results for the bright sources Crab Nebula, GS2023+338 and MAXI J1820+070\footnote{Based on observations with \textit{INTEGRAL}, an ESA project with instruments and science data centre funded by ESA member states (especially the PI countries: Denmark, France, Germany, Italy, Spain, and Switzerland), Czech Republic and Poland with participation of Russia and USA.}}

\begin{abstract}
{The \textit{INTEGRAL} SPI instrument observes the hard X-ray sky from 20 keV up to a few MeV since more than 15 years. In this energy domain, the main emitters are compact objects for which SPI provides spectral information of prime interest. Recently, two transient sources reached very unusual flux levels and have been detected up to a few hundreds of keV with a high significance level. A drastic reduction of the systematic errors is thus required to obtain reliable spectra. This objective is achieved through an analysis including a detailed understanding of the instrument behavior. This paper presents both aspects of the data analysis:  we first give a basic description of the instrumental issues, then we present the solution to be implemented in the SPI data analysis (at the event selection stage) and illustrate with a few examples the reliability of the SPI results in the high-energy domain when the data analysis is performed properly. We take benefit from this refined analysis procedure to propose an updated model of the hard X-ray spectral shape of the Crab Nebula. We revisit the high-energy emission observed in GS2023+338 spectra during its 2015 outburst and present the first results from the SPI observations dedicated to the recently discovered  transient MAXI J1820+070.}
\end{abstract}
\keywords{X-rays: individual (Crab Nebula, GS2023+338, MAXI J1820+070);  X-rays: binaries; gamma-rays: general}
\section{Introduction}
SPI is a hard X-ray/gamma-ray telescope providing an excellent energy resolution in the 20 keV- 8 MeV energy range. This is one of the main instruments of the \textit{INTEGRAL} ESA mission launched on October 17, 2002.  SPI is described in \cite{Vedrenne03} and a first view of its in flight performance can be found in \cite{Roques03}. As a spectrometer, SPI has to provide precise and reliable spectra of the observed celestial objects. To that end, a solid work of calibration has been performed with extensive ground calibrations \citep{att2003} and, in parallel, an important simulation work \citep{sturner2003}. We consider SPI calibration as absolute, and we have measured a spectrum from Crab Nebula \citep{Crab09} which is considered as a standard candle in this energy domain.

Measurements of high-energy source spectra are a key to understand and disentangle the processes at work during the emission of hard X-ray/gamma-ray photons. For X-ray binaries, the high-energy emission allows exploring the hot corona configuration, the  jet contribution (synchrotron emission and its potential polarization) as well as the interactions between the various mechanisms of energy transfer. For instance, several sources exhibit a hard component, beyond the Comptonization emission (i.e. above ~200 keV), bright enough to be precisely studied with hard X-ray instruments such as \textit{INTEGRAL} SPI. 
Recently (in 2015 and 2018), two objects have presented intense activity periods, with maximum fluxes up to 40 Crab (GS 2023+338 = V404 Cygni) and 4 Crab (MAXI J1820+070). This has been the opportunity to improve our knowledge of some instrumental effects and to refine our data analysis procedure. 
In this paper, we aim to explicit our approach and share the results of our comprehensive study.
This requires some knowledge of the SPI event tagging mechanism and the understanding of the defects affecting the pulse processing. 
Thus, in section 2, we describe the on-board event processing together with the spectral characteristics of the different event families. We study the systematic errors that can appear according to the event selection, and show how simple recommendations provide clean spectra.
In section 3, we apply this work to scientific purposes. We first analyze two datasets dedicated to the Crab Nebula. Then, we revisit some observations performed during the exceptionally bright outburst of the X-ray transient GS 2023+338  and finally, we present the first results on the high-energy emission of  the recently discovered transient MAXI J1820+070.

\section{Understanding instrumental systematic errors}
%%%%%%%%%%%%%%%%%%%%%%%%%%%%%%%%%%%%%%%%%%%%%%%%%%%%%%%%%%%%%%%%%%%%%%%%%
\subsection{Description of SPI on-board pulse processing}
%%%%%%%%%%%%%%%%%%%%%%%%%%%%%%%%%%%%%%%%%%%%%%%%%%%%%%%%%%%%%%%%%%%%%%%%%
This section outlines a synthetic view of the on-board processing of the
detection plane events. Both analog and digital processing will be discussed, focussing on the main characteristics relevant to understand the results presented below.
\paragraph{GeDs AND PREAMPLIFIERS}

The SPI detection plane is composed of 19 High Purity Germanium detectors (GeD) cooled at 80 K.
Each Germanium central contact (anode) is AC coupled to a preamplifier cooled at -65 $^\circ$C.
Each preamplifier outputs two signals:
One is the "energy" signal, which is the classical output for a charge sensitive preamplifier. This output pulse is "charge integrated" and its amplitude is proportional to the energy deposit. Each pulse has a decay time constant of 500$\mu$s.
The second output is the "PSD" signal (PSD for Pulse Shape Discrimination), which represents the evolution of the current delivered by the Germanium during an interaction. The PSD output pulses have a maximum duration of a few hundreds of ns.

\paragraph{THE ANALOG FRONT-END ELECTRONICS (AFEE)}
Each energy signal (from each of the 19 preamplifiers) feeds into a dedicated AFEE channel.
An AFEE channel contains: a pulse shaping electronics, a pulse height analyzer, an ADC (analog to digital converter), a configurable low energy threshold (LLD) and a fixed ($\approx$ 8 MeV) upper level threshold (ULD).  It delivers the energy of the incoming signal together with a time tag that corresponds to the time of the LLD trigger. After an LLD trigger for an event below the ULD, the electronics module is occupied (dead time) during 26$\mu$s. In case of an event above the ULD, a saturating event time tag is issued, and the electronics module is blocked for 100 $\mu$s.
\paragraph{THE PULSE SHAPE DISCRIMINATION ELECTRONICS (PSD)}
  The primary goal of the PSD electronics device was the analysis of the Germanium's pulse shape for event discrimination.
The PSD electronics feature 19 independent configurable LLD's and ULD's, a fast digitization of the input pulse and its analysis against a library. An internal coincidence mechanism rejects multiple detector events, i.e. events that trigger more than one PSD LLD in the same time window.
As output, among other information, the PSD electronics deliver a Time Tag, for each triggered event, i.e. for an event between LLD and ULD on only one detector.
Even with a very fast front end trigger processing, the PSD has its own dead time which leads to a PSD efficiency that will be discussed later on. This dead time is determined by the counting rates incoming in the PSD electronics and the PSD front end configuration setup (threshold values).

\paragraph{THE DIGITAL FROND END ELECTRONICS (DFEE)}
The DFEE analyzes the events detected by the 19 AFEE electronic chains, using the different received time tags. Simultaneously, the precise dating of each event is realized. All this is done in real time.
This process runs as follows:
\begin{itemize}
\item All the time tags are aligned together, taking into account the various hardware delays.
\item Coincidence time windows are created in order to find:
\begin{itemize}
\item Potential association between events localized in different detectors, using AFEE time tags: creation of Multiple Event block (ME).
\item Potential association between an AFEE time tag and a PSD time tag: detector event analyzed by the AFEE and triggering the PSD: creation of a PSD event (PE)
\item Potential association between an event and an anti-coincidence (ACS) time tag then creation of vetoed/non-vetoed events.
\end{itemize}
\item After this "association machine", the events are classified as follows, in the nominal configuration of SPI for science operations:

\begin{itemize}
\item The events in coincidence with an ACS time tag (vetoed events) are "dropped" from science usage. Otherwise, each event belongs to one and only one of the following categories:
\item  SE = non-vetoed event, non-coincident with anything else. It corresponds to one AFEE trigger.
\item  ME = two or more non-vetoed events  falling in the same time window.  It corresponds to two OR more AFEE triggers.
\item  PE = non-vetoed event falling in the same time window than a PSD time tag. Hence, it corresponds to one AFEE trigger AND one PSD trigger.
\end{itemize}
\item The DFEE electronics also measure the effective dead time that applies to the non-vetoed events coming out of each AFEE electronic chain. This dead time takes into account the combination of the electronic chain blocking durations.

\end{itemize}

\subsection{Spectral characteristics of the event categories}
\begin{figure}
\includegraphics[scale=0.4]{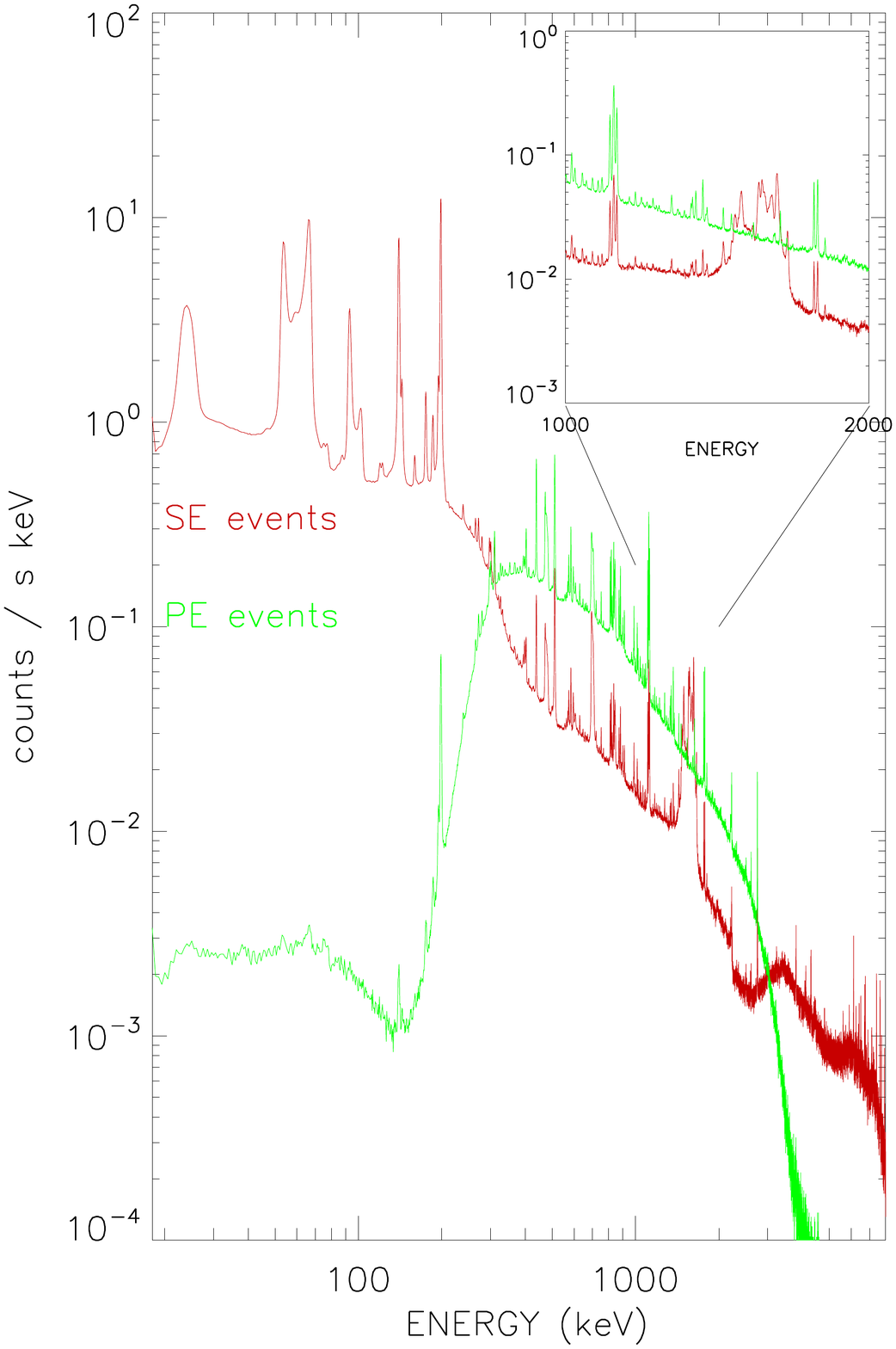}
\caption{Example of SPI background spectra using SE (red curve) and PE (green curve) datasets. The insert points out the 1.5 MeV feature.}\label{bdfpese}
\end{figure}

In this section, we study the characteristics of the spectra obtained from the on-board classification of the events. We discuss the links between the physics of the photon interactions in the SPI detection plane, the electronics behavior, the triggering scheme and the event classification.
The spectra and the results presented below concern the whole detection plane. It means that the spectra are constructed regardless the detector number.
Figure \ref{bdfpese} displays spectra built using the SE events only (in red) and  the PE events only (in green), from 20 keV (AFEE LLD) to 8.0 MeV (AFEE ULD).
\paragraph{The Single Events (SE)}
From the triggering scheme point of view, an SE event corresponds to a unique AFEE time tag, not coincident with anything else.
In figure \ref{bdfpese}, the SE spectrum exhibits a deficit of counts in the $\approx 200-3000$ keV range: in this range, the events classified as PE are missing. In addition, a feature around 1.5 MeV is obvious. This feature, originally called "electronic noise", is known since the \textit{INTEGRAL} launch, and first reports can be found in \cite{Wund2004} and \cite{Wund2005}. During data analysis, attention must be paid to avoid any impact on the source flux extraction. We pointed out this issue in \cite{Crab09}, along with a first recipe to solve it.

\paragraph{The Multiple Events (ME)}

A Multiple Event block corresponds to two or more AFEE time tags occurring in the same time window and non-vetoed (outside an ACS time window). Physically, this requires that at least two photons interact with two different germanium detectors, at the same time.
Typically, it happens when a photon interacts by Compton effect with one GeD, and the diffused photon interacts with another detector.
The dead time corresponding to ME is higher than for the single interactions, because two (for a double event) or more AFEE electronic chains have to be available.
We have computed this additional dead time for double events; it ranges from 4.6 \% (low background) to 7.5 \% (high background).
This means that a fraction of the double interactions (4.6 to 7.5 \%) is classified as single interactions, associated with only a part of the total energy deposit. The same reasoning can be made for triple events (a potential ME3 becomes ME2) and so on.

\paragraph{The Pulse Shape Discrimination events (PE)}

An event is classified as PE when a PSD time tag occurs within an AFEE time window for the same detector. It should be noted, and this is of prime importance for  the later discussion, that this requires two triggers from two different types of analog pulse analysis from two different outputs from the same preamplifier.
In Figure \ref{bdfpese}, the PE spectrum shape (green curve) illustrates the behavior of the PSD electronic function.  It selects events in the range $\approx 200-3000$ keV, with a maximum efficiency between 400 and 2200 keV. It is important to note that threshold values and full efficiency depend on the PSD front-end configuration parameters. These parameters have been changed on a few occasions over the \textit{INTEGRAL} mission and are known at anytime, from the telemetry data (see Table \ref{PSDth}).

\paragraph{Efficiency of the PSD electronics} 

The PSD electronics device has its own dead time, which limits its efficiency. The PSD efficiency corresponds to the number of PSD tagged events divided by the sum of the single detector interaction events (PE+SE). We can measure this efficiency using gamma-ray lines, produced by internal radioactive elements and visible in the SPI background spectra.
Using gamma-ray lines ensures that: i) the photons have a physical origin; ii) their energy is properly measured; iii) we get the total energy deposit of the photon (no multiple detector interactions). We exclude multiple photon radioactive decays, as well as 511 keV photons, that may be part of an ME.

Table \ref{tabPSD} provides the PSD efficiency measured for a few revolutions with a set of background lines: \added{ 271.26 keV from ${}^{44m}\text{Sc}$,} 309.87 keV from ${}^{67}\text{Ga}$, 351.93 keV from ${}^{214}\text{Pb}$, 403.19 keV from ${}^{67}\text{Ga}$,\added{ 438.62 keV from ${}^{69m}\text{Zn}$,} 584.54 keV from ${}^{69}\text{Ge}$, 1117.26 keV from ${}^{69}\text{Ge}$,\added{ 1764.34 keV from ${}^{205}\text{Bi}$,} 2223.25 keV from ${}^{1}\text{H}$ and 2754.0 keV from ${}^{24}\text{Na}$.

If we exclude the spectral regions affected by the thresholds, the PSD efficiency ($eff_{psd}$) is remarkably stable and ranges from 0.88 \deleted{\%} in revolution 43 (March 2003) to 0.85 \deleted{\% }in revolution 1856 (March 2017), thus over the course of 15 years of operation.  We also demonstrate that this efficiency can be precisely measured at any time.

In summary of this study, we have shown that SE events produced by the SPI tagging system are clearly affected by spurious artefacts with a particular enhancement in the MeV region, while, thanks to a more complex triggering scheme, the PE events are unaffected.

\subsection{The origin of spurious events}
From ground tests, we know that the spurious events in the SE spectra are linked to high-energy deposits in the Germanium detector. These events saturate the analog electronics and, due the large peaking time of the shaper (8 $\mu$s), the recovery time of the base line for energy measurement is very long. Thus, the energy determination of events falling during the recovery time is wrong and pushed towards higher energies.
Conversely, the PSD trigger electronics is not affected by this problem: it uses the current pulse output of the preamplifier (and not the charge integrated output), with a very short integration time (40 ns).
Hence, when the PSD electronics produce a time tag, the event energy is actually in the range delimited by the LLD and ULD of the PSD. Consequently, the use of the PSD time tag allows a confirmation of the event energy range and ensures that it is not a low energy event shifted to higher energy.

\subsection{Source spectra and event selection}
\begin{figure}
\includegraphics[scale=0.4]{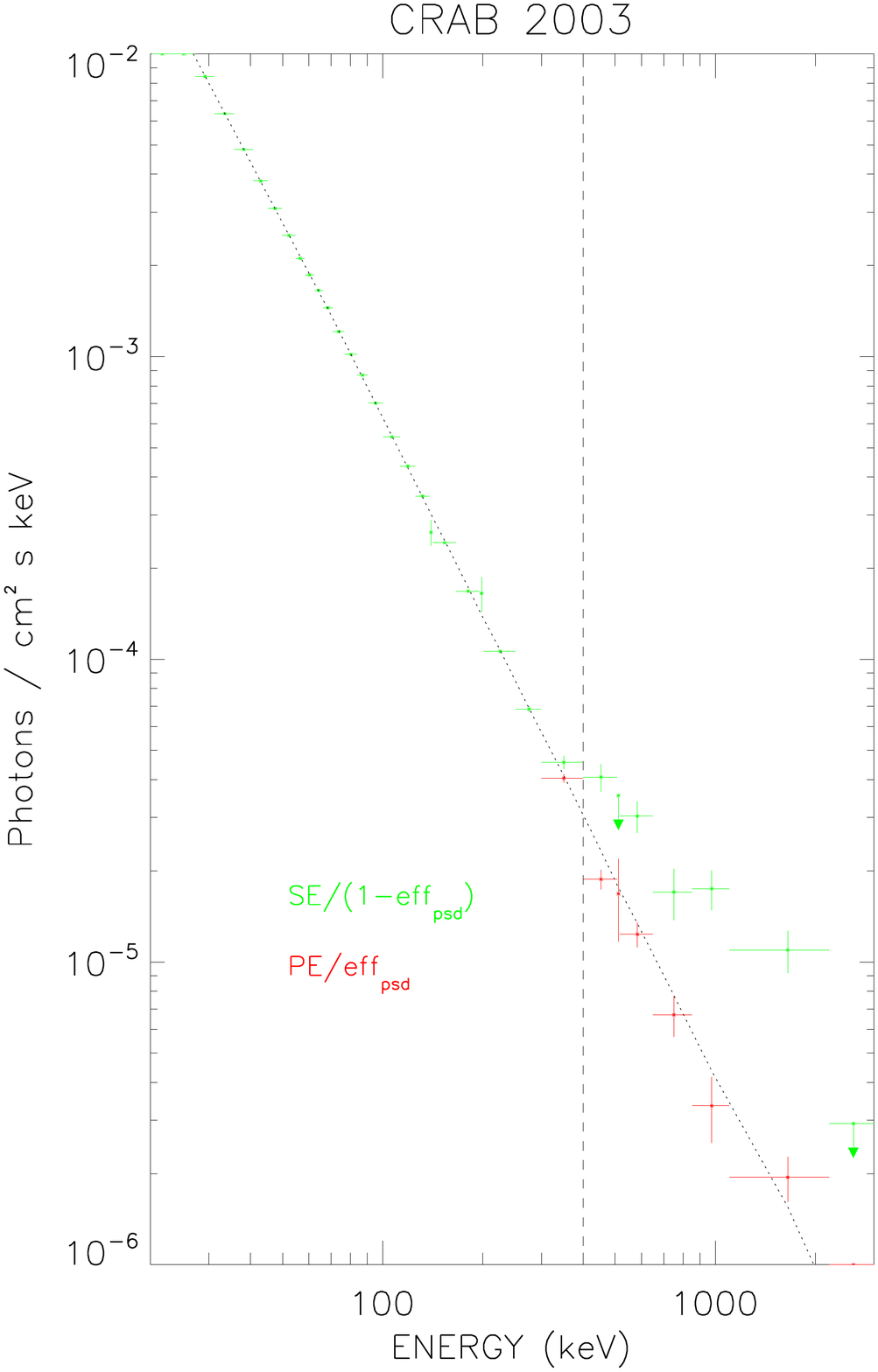} 
\includegraphics[scale=0.4]{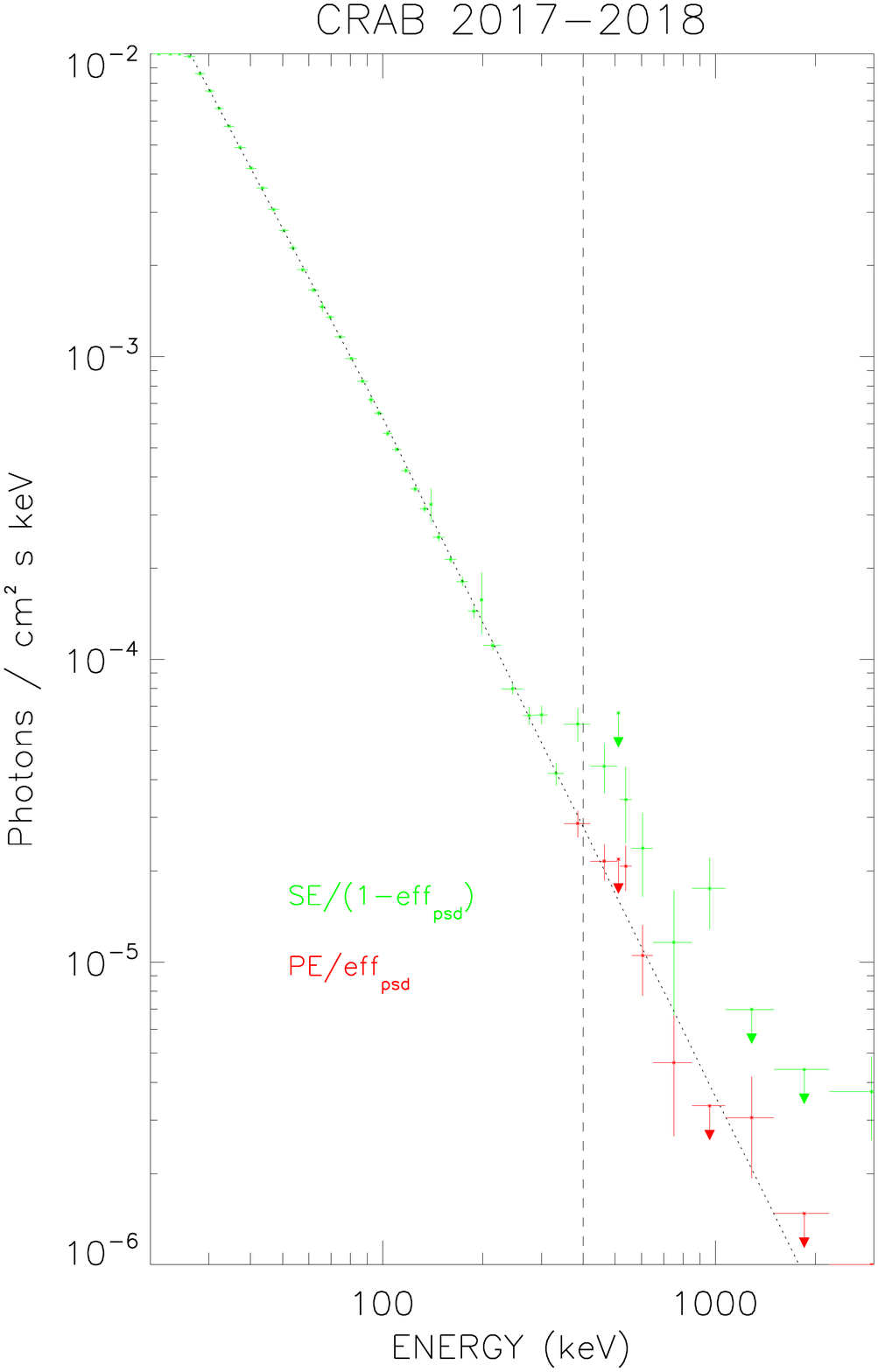}
\caption{Crab Nebula spectra obtained with SE and PE datasets for 2003 (left panel) and 2017-2018 (right panel) observations.}\label{crabpese}
\end{figure}

As a next step, we determine how the source flux extraction is affected by the phenomenon described above. We have chosen two bright hard X-ray emitters to quantify the fraction of photons flagged by the PSD electronics (PE events) relatively to the total number of photons attributed to the source, after the deconvolution process (i.e. taking into account the imaging information).
The data have been processed using our usual analysis pipeline but splitting the incident counts into two datasets, according to the value of their PSD flag. We have then performed the flux extractions separately and built the spectra with the associated matrices, for each datasets. If everything works properly, for E between 400 keV and 2700 keV, count (and then photon) spectra should follow the simple relations:

\begin{displaymath}
C_{PE}/C_{tot}=eff_{psd},
C_{SE}/C_{tot}=1-eff_{psd}
\end{displaymath}

where $eff_{psd}$ represents the PSD efficiency studied in section 2.2.
In other words, we should retrieve the same source spectrum both from SE counts and PE counts, when applying the appropriate normalization factors.
We first test  this procedure with Crab observations covering revolutions 43-44-45 (beginning of the mission; March 2003). In figure \ref{crabpese} (left panel), are displayed the photon spectra extracted from PE dataset and SE dataset, renormalized by the factor corresponding to the PSD chain deadtime. This latter has been determined using background lines, as previously described, and we get $eff_{psd} =0.88\deleted{\%}$, (see Table \ref{tabPSD}) .\\
It appears that the renormalized spectra disagree above  $\approx$ 400  keV, with the reconstructed SE spectrum containing significantly more photons than that extracted from PE events.\\

\begin{figure}
\includegraphics[scale=0.4]{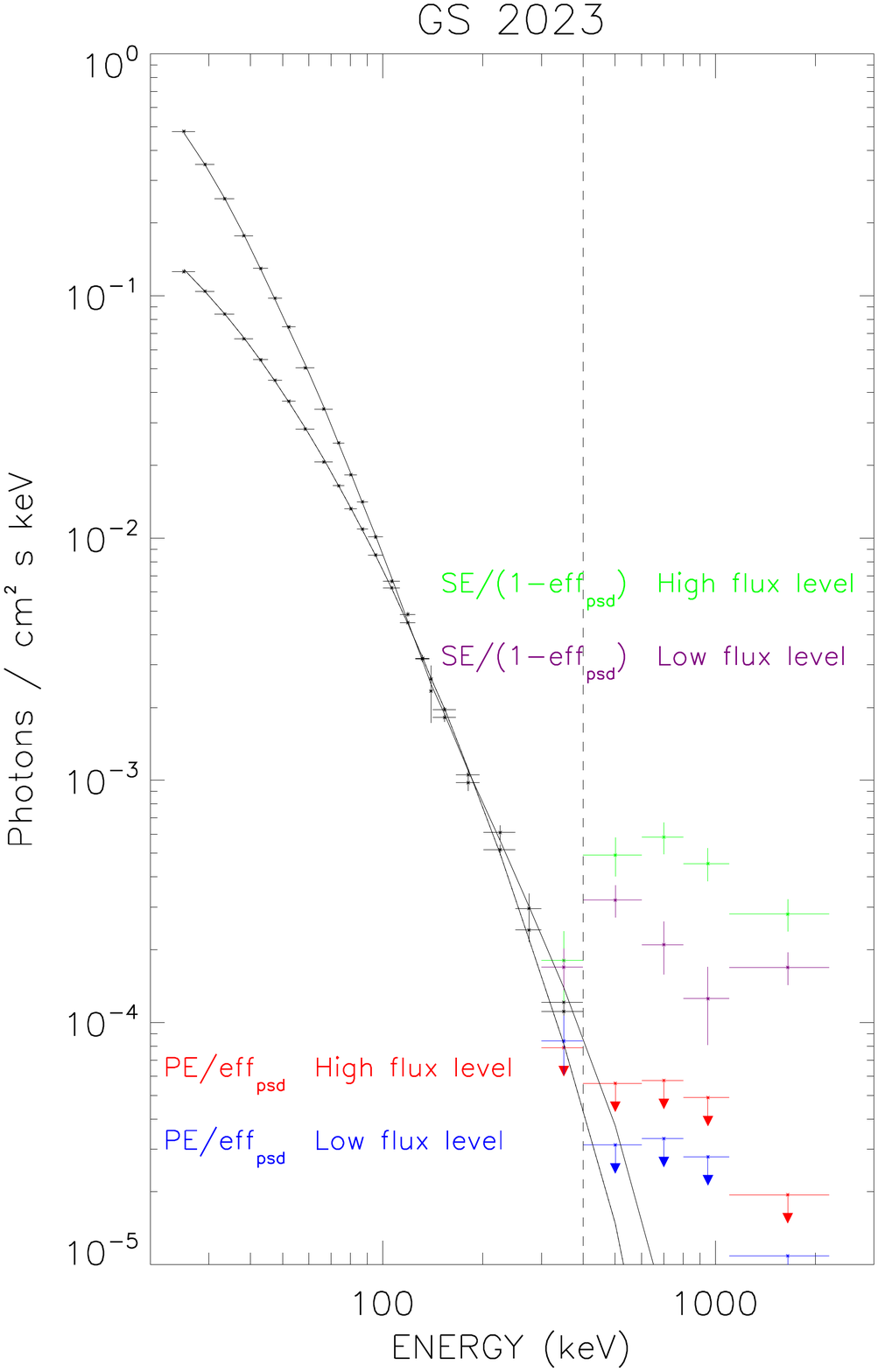} 
\caption{GS2023+338 spectra for SE and PE datasets for two flux levels.}\label{V404pese}
\end{figure}

We have repeated the procedure for later Crab observations (Revolutions 1856-1857+1927-28; August 2017 and March 2018), for which $eff_{psd} =0.85\deleted{\%}$.
Here too (see figure \ref{crabpese}, right panel), the spectrum based on SE events deviates from the one based on PE events, with higher fluxes above $\approx$ 400 keV.\\
To get higher statistics, we have then considered an observation of the very bright transient GS2023+338 during its 2015 June outburst. We have chosen two 1 hour exposures presenting two different flux levels during revolution 1557. We extracted spectra from SE and PE events respectively, for both exposures. The four spectra are presented in figure \ref{V404pese}, revealing the discrepancies between SE and PE datasets, for each exposure.
Comparing more precisely the resulting fluxes versus the source intensity points out that the "false" SE rate above 400-500 keV increases with the (low energy) source flux. This supports the hypothesis that the 'false' single events above 400 keV are low energy photons, shifted at higher energy due to a pile-up on the residual baseline in the electronic chain, after saturating events.
This correlation also demonstrates that the shifted photons come from the source. Indeed, if the observed effect were due to background photons, it would not depend on the source flux. Moreover, background photons are eliminated during the image "de-convolution" process. Another implication is that, in practice, the effect is significant for bright sources only (above $\sim$ 1 Crab).

\paragraph{The fraction of "false" events in single event spectra}
The last step of this study is to estimate the fraction of events whose energy has been corrupted. To that end, we have to look at the difference between the spectra reconstructed with the PE events and the spectra reconstructed with the sum of the PE and SE events.  We then compare the excess photon flux  in the later to the source flux at 60 keV (representative of the source low energy flux). 
This ratio has been computed for source intensities ranging from 1 to 40 Crab. We found that the flux of "false" SE events in the 350-2000 keV energy range represents 0.12\% to 0.20\% of the 60 keV flux (expressed in $ph~cm^{-2} ~s^{-1})$. This fraction depends on the saturating event flux and on the source spectral shape at low energy. 

\subsection{Instrumental study conclusions}

From the investigations presented above, we conclude that the primary electronic chain (AFEE) of the SPI detector plane displaces a tiny fraction of the analyzed photons toward higher energies. Because celestial sources emit many more photons at low energy than at high energy, even if only a small fraction of low energy photons are shifted, their number may be commensurable to the photons produced above $\approx$ 400 keV. It becomes negligible as soon as the source flux is more important. Moreover, this effect increases with the source intensity and must be taken into account for bright sources (1 Crab for a typical observation of one revolution).  However, we have shown that, due to a more complex triggering scheme, the PE events are not affected by this electronic problem. Thus, the obvious solution is to use the PE events whenever this is possible. Further details on the recommendations for a reliable data analysis can be found in Appendix B.

\section{On the high-energy emission of three bright hard X-ray sources}

We apply the procedure described above to the spectral analysis of three among the brightest hard X-ray emitters: the Crab Nebula, GS2023+338 (which reaches intensity levels of  30-40 Crab during its 2015  outburst) and  MAXI J1820+070 (which reaches an intensity of 4 Crab during its 2018 outburst).

\subsection{Crab Nebula}
The Crab Nebula is the brightest persistent source in the hard X-ray sky and often used as a reference in this domain.
Its stability in shape as well as in intensity with time (except some flares recurrently observed in the GeV/TeV energy range) makes it possible to handle its global spectral emission from radio to Very High Energy, even from non-simultaneous observations. \deleted{A long-term study by \cite{WH11} has revealed slight but significant variations of the flux (of the order of $\pm$ 1 to 2 \% per year), plus a more important evolution  (decrease by 7-8 \% between 2008 and 2010), affecting the Nebula emission. While these variations may undermine the "standard candle" status of this source,  their amplitudes remain very low and slow around a steady mean, keeping it as the best reliable candidate for instrument cross-calibration and checking.}

In a previous paper \citep{Crab09} reporting on the \textit{INTEGRAL} SPI observations, we compared spectra obtained during several periods along the first 6 years of the \textit{INTEGRAL} mission, demonstrating the stability of  both the SPI spectrometer and the Nebula spectral emission, within  $\approx 5\%$ of uncertainty. 
Here, we apply our refined analysis to provide an updated determination of the hard X-ray/soft gamma-ray emission as measured by SPI. 
We have used two datasets, one corresponding to the observations performed during the first year of the \textit{INTEGRAL} mission and the second encompassing ten revolutions performed between 2016 and 2018.

\begin{figure}
 \includegraphics[scale=0.45]{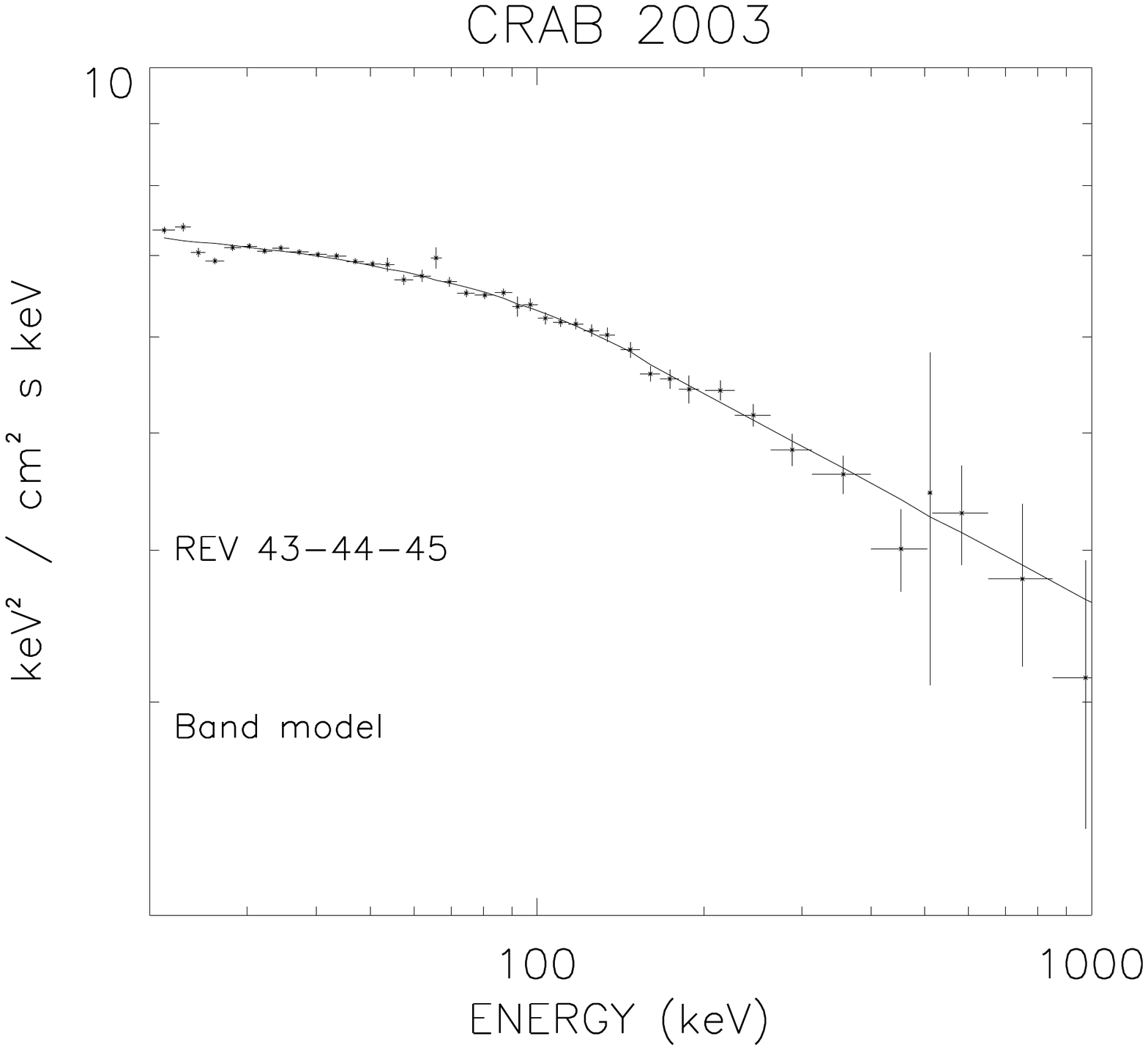} 
 \includegraphics[scale=0.45]{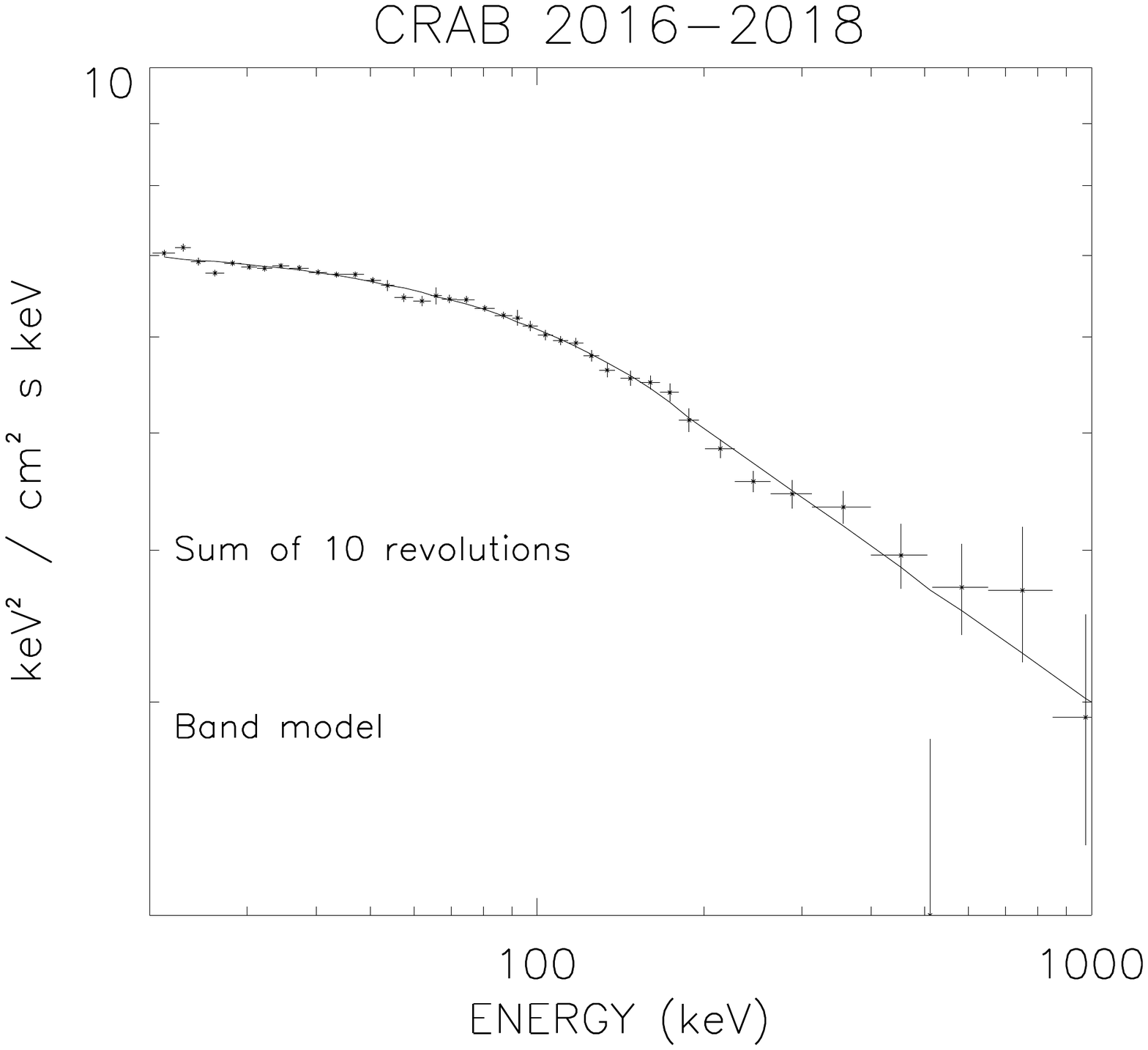}
 \caption{Crab Nebula spectra from 2003 and 2016-2018 observations.}\label{figCrab}
\end{figure}

\begin{figure}
\includegraphics[scale=.6]{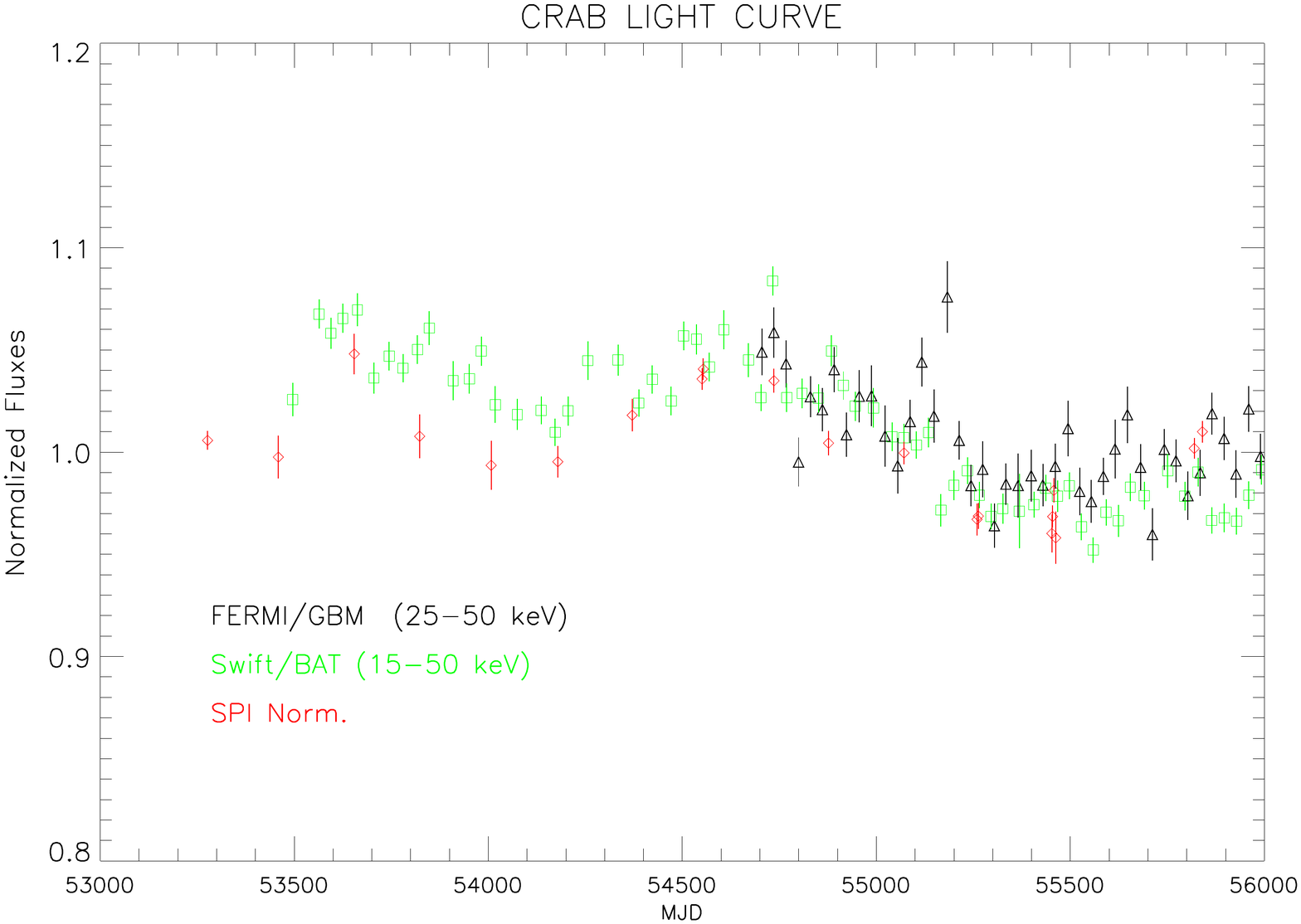}
\caption{Crab Nebula flux evolution, with statistical errors only. \textit{Fermi} GBM Earth occultation results provided by the \textit{Fermi} GBM Earth occultation Guest Investigation teams at NASA/MSFC and LSU. \textit{Swift}/BAT transient monitor results are provided by the \textit{Swift}/BAT team.}\label{CrabLC}
\end{figure}

While the spectral emission is generally well described by a power law in any limited energy range (typically instrument energy ranges), we had pointed out this is not the case in the SPI energy range. Using a broken power law model, the spectral slope evolves from a photon index of 2 to an index of 2.2-2.3 in the 20-1000 keV domain. The energy break had been fixed (to 100 keV), to avoid degeneracy between the parameters and allow an easy comparison of the spectral shape between observations. However, a smooth curvature provides a better description of the data. A continuously varying slope model has been proposed by \cite{Massaro}.
\begin{displaymath}
F(E) = A.E^{a+bLog(E/E_0)}.
\end{displaymath}
 Unfortunately, this model cannot be extrapolated outside of the considered energy range, since the slope evolves rapidly toward very hard (at low energy) or very soft (at high-energy) values. We opt for a Band model proposed by \cite{band93} to describe GRB spectral emission, which joins two power laws by a smooth curvature:
\replaced{
\begin{displaymath} A(E)= \left\{
\begin{array}{ll}
A(E)=K(E/100)^{\alpha_1} \exp(-E/E_c)   &  if E > E_c(\alpha_1 - \alpha_2) \\
A(E)=K((\alpha_1-\alpha_2)E/100)^{(\alpha1-\alpha_2)} \exp(-\alpha_1 - \alpha_2 ) &  if E < E_c(\alpha_1 - \alpha_2) 
\end{array} \right.
\end{displaymath}
}
{
\begin{displaymath} A(E)= \left\{
\begin{array}{ll}
A(E)=K(E/100)^{\alpha_1} \exp(-E/E_c)   &  if E < E_c(\alpha_1 - \alpha_2) \\
A(E)=K[(\alpha_1-\alpha_2)E_c/100]^{(\alpha1-\alpha_2)}(E/100)^{\alpha_2} \exp(-(\alpha_1 - \alpha_2) ) &  if E > E_c(\alpha_1 - \alpha_2) 
\end{array} \right.
\end{displaymath}
}
where $\alpha_1$ and $\alpha_2$ are respectively the low and high-energy power law slopes, and $E_c$ is a characteristic energy. \\
The best-fit parameters are given in table \ref{tabCrab}, while both spectra are displayed on figure \ref{figCrab}. The resulting $\chi_{red}^2$ values for the Band model are 1.9 and 1.36, to compare to 2.32  and 1.87 for a broken powerlaw model with a free break energy. \added{To minimize the fit degeneracy between slopes and $E_c$ values, we have fixed $E_c$ to 500 keV. The shape parameters remain very similar. Note that the global normalization is slighlty  decreasing, by less than 5\%}.
\added{A long-term study by \cite{WH11} has revealed slight but significant variations of the flux (of the order of $\pm$ 1 to 2 \% per year), plus a more important evolution  (decrease by 7-8 \% between 2008 and 2010), affecting the Nebula emission. This result is confirmed by the SPI observations. We have fit the SPI spectra obtained during the same period with a Band model ($E_c$ fixed to 500 keV) and use the model normalization as the tracer of the source intensity. In fig. \ref{CrabLC}, we compare the evolution of this parameter with the Crab Nebula light curves obtained from \textit{FERMI} GBM and  \textit{Swift} BAT  (averaged over ~ 1 month) to illustrate the good agreement between instruments. While these variations may undermine the "standard candle" status of this source,  their amplitudes remain very low and slow around a steady mean, keeping it as the best reliable candidate for instrument cross-calibration and checking.}

In conclusion, the curved spectral shape provides an appropriate description of the Crab Nebula emission, improving substantially the $\chi^2$ values relatively to a broken power law model in the hard X-ray domain. It could be used to fit the source emission from X-rays up to MeV region or more, in particular for cross calibration purposes. 
Lastly, the Crab Nebula emission appears very stable in shape and in intensity for these two periods 15 years apart.

\subsection{GS2023+338 high-energy emission}

\begin{figure}
\includegraphics[scale=0.9]{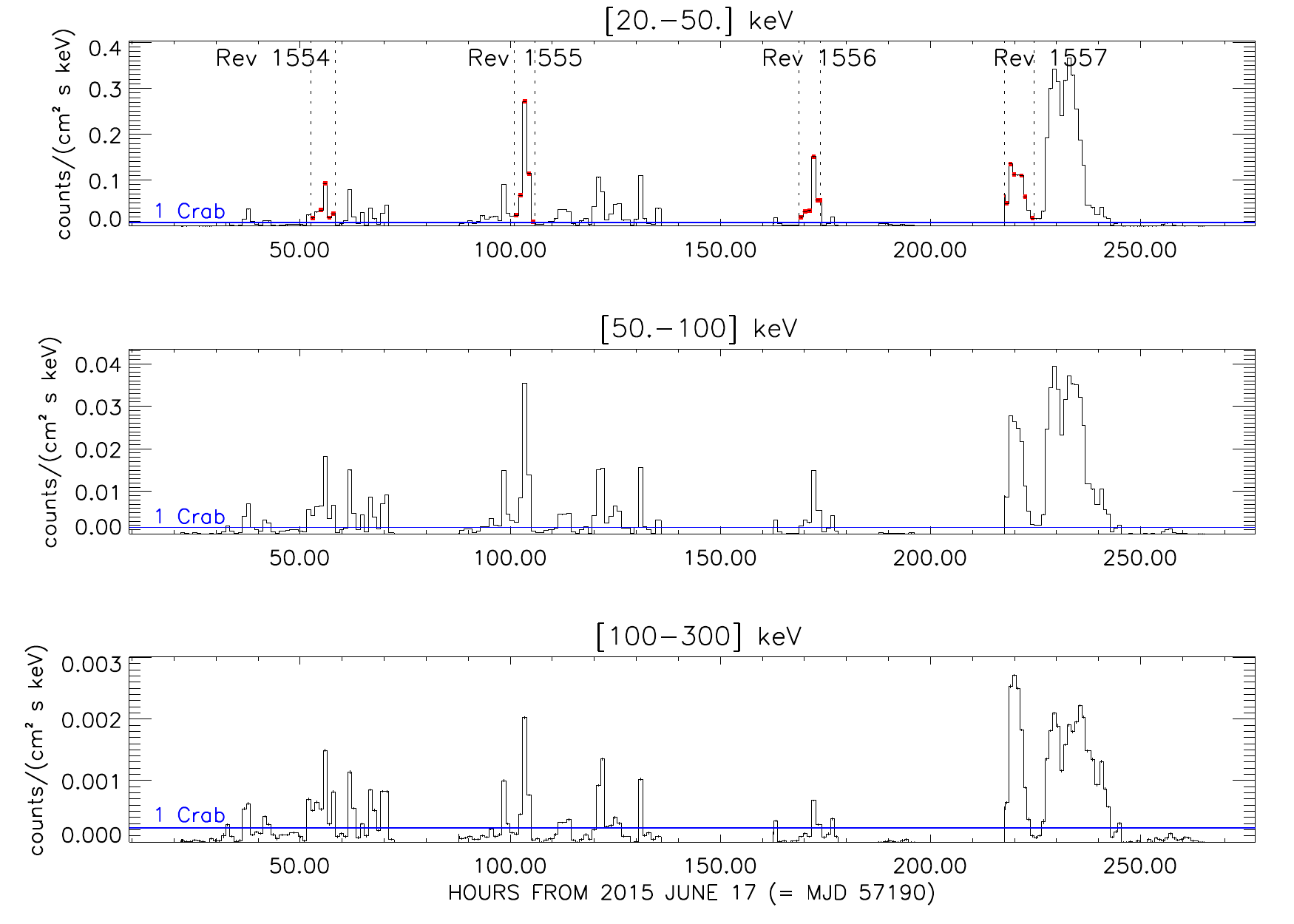} 
\includegraphics[scale=0.9]{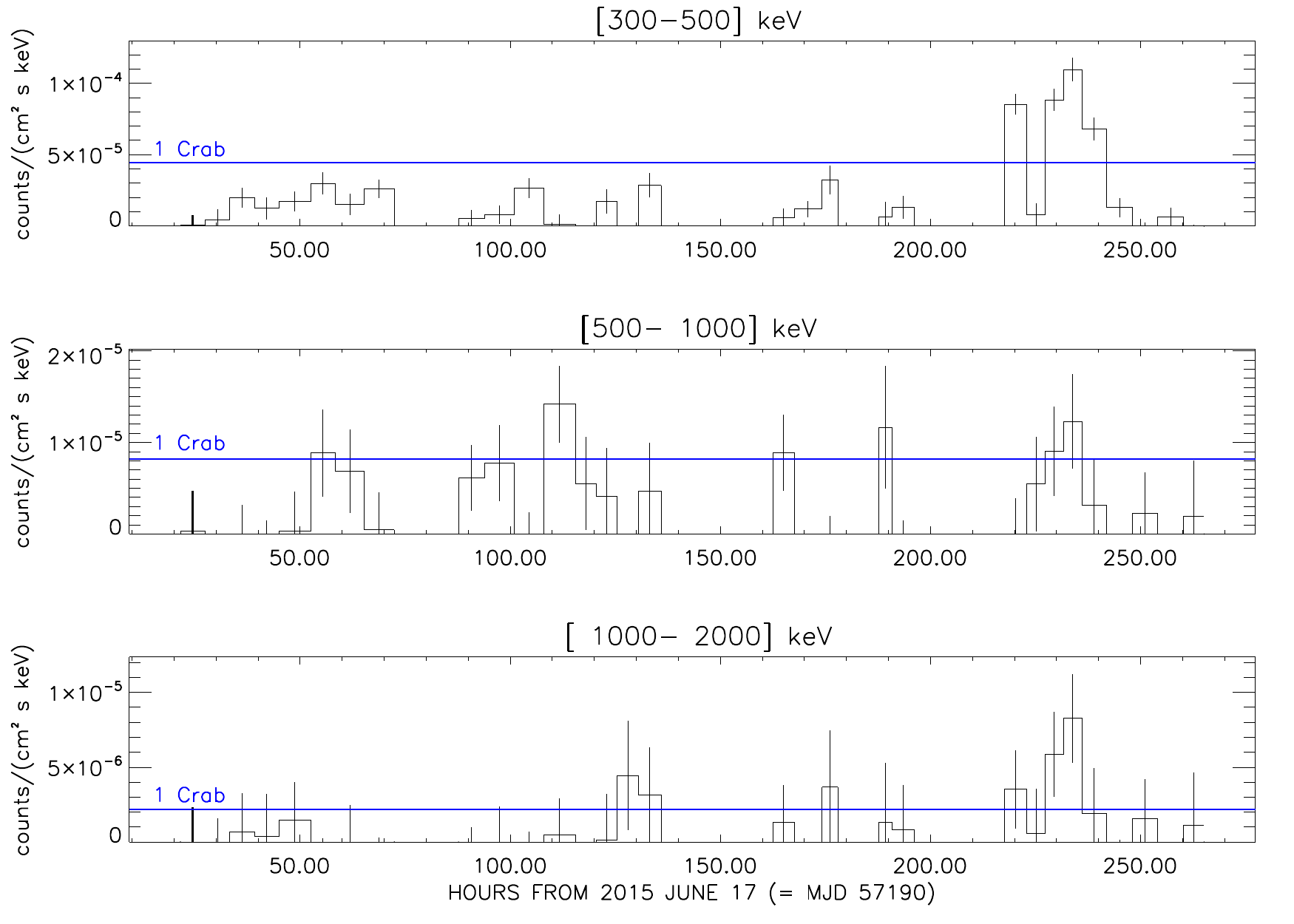}
\caption{Light curves observed for GS2023+338 by \textit{INTEGRAL} SPI in six energy bands, from the 15th to the 27th of June, 2015. The timescale is 1 scw (i.e. $\approx$ 1 hour) for the three low energy bands and 6 scw for the three high-energy bands.}\label{LCV404} 
\end{figure}

The \textit{INTEGRAL} mission provided a large coverage of the GS2023+338(=V404 Cygni) 2015 summer outburst.
The most active part of the outburst occurred during revolutions 1554 to 1557 (see \textit{INTEGRAL} observation information in Table \ref{tabV404}). Fig. \ref{LCV404} displays the flux evolution of GS2023+338 on a $\approx$1 hr timescale for energies from 20 to 300 keV, and a $\approx$ 6 hour timescale from 300 keV to \replaced{1} {2} MeV for statistical reasons. Beyond the huge variability of the source, we can see that the flux evolution is energy dependent.  In this paper, we focus on the high-energy emission of this impressive transient source, by comparing the source behavior during four flaring episodes, identified by vertical dashed lines in Fig. \ref{LCV404}.

We have built individual spectra along each of the considered flares and fit them within Xspec. 
Two components are needed to describe the spectral emission of GS2023+338 between 20 keV and $\approx$1 MeV, as for Cygnus X-1 and other bright hard X-ray emitters.
For the low energy part (between 20 and 100-150 keV), several scenarios can be assumed. Usually, one assume a Comptonization of low energy photons (from an accretion disk) by a hot electron population in thermal equilibrium, with a reflection component. However, GS2023+338 presents a unusual curvature below 30-50 keV, which can be reproduced allowing either a characteristic temperature of 5-7 keV for the seed photon population (in this case, it is probably produced by synchrotron emission) or a strong absorption, with typical column density of a few $10^{23}cm^{-2}$. We also notice that this curvature makes the low energy spectral shape resembling to a reflection dominated (or pure reflection) emission. Since it is impossible to discriminate between all these possibilities, we have investigated two extreme scenarios to account for the first component: the first one relies on an absorbed Comptonization component (seed photon temperature fixed to 0.3 keV), without reflection, to limit the number of free parameters, while in a second step, we will consider a pure reflected emission.  \\
The second component extends up to $\approx$800 keV. We modeled it by a cut-off power law, with a photon index of 1.6 and an energy cutoff of 220 keV. Due to the lack of statistics, we let the normalization as the only free parameter. The value of the power law index could interplay with the Comptonization parameters (kT and $\tau$). However, we have checked that the global conclusions of our study remain unaffected. As a last assumption in our approach, we have favored scenarios where the source configuration presents the highest stability. In this idea, for each flare, we have fitted simultaneously all the spectra obtained on the one scw ($\sim $ one hour) timescale, and sought for the minimum of free parameters. We thus started with the same parameter values (except all normalization factors) for the given set of spectra, and then freed them one by one, when  justified by the $\chi^2$ improvement. 

\paragraph{Highly absorbed Comptonization emission model}
\begin{figure}
 \includegraphics[scale=0.4]{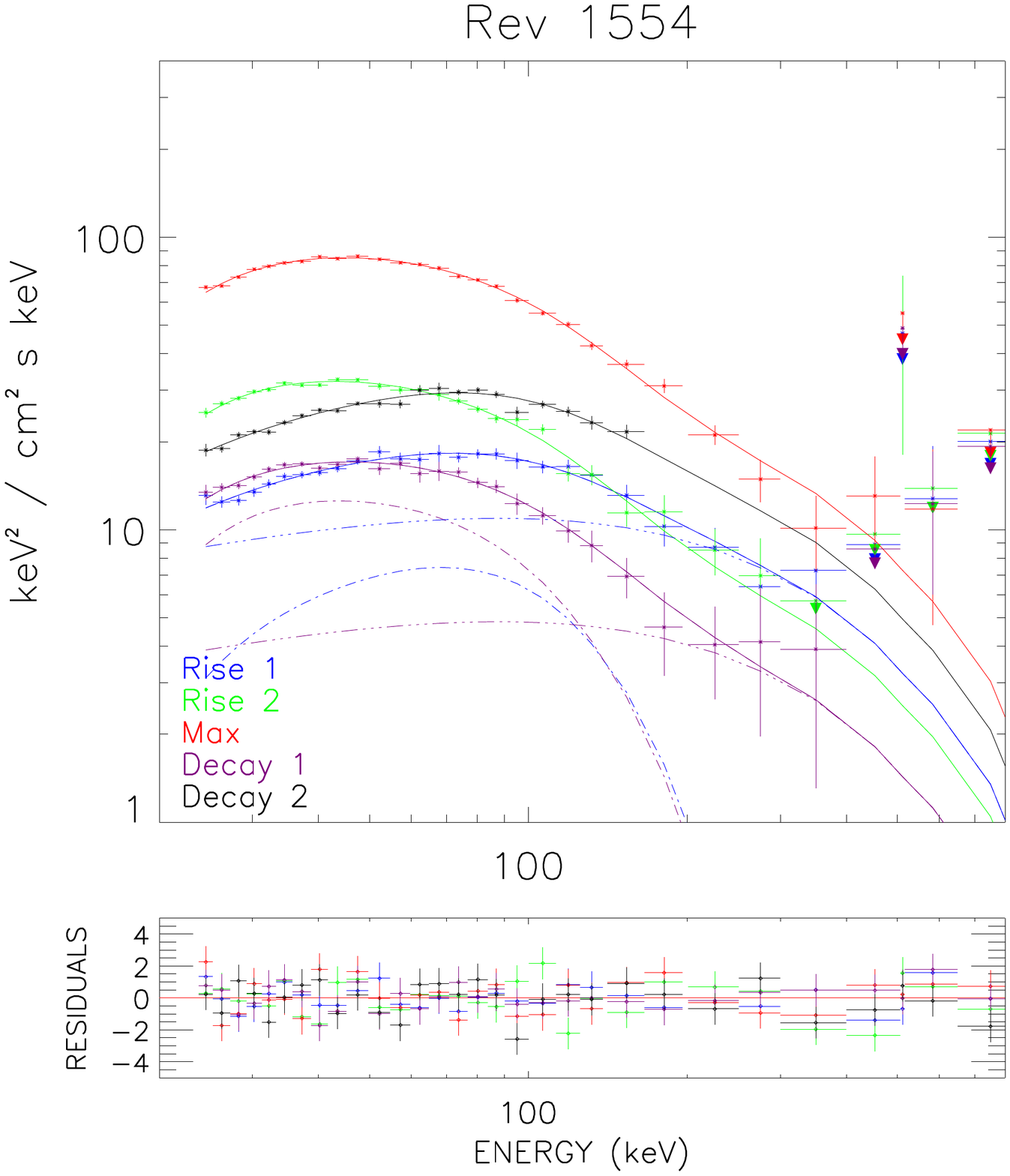} 
 \includegraphics[scale=0.4]{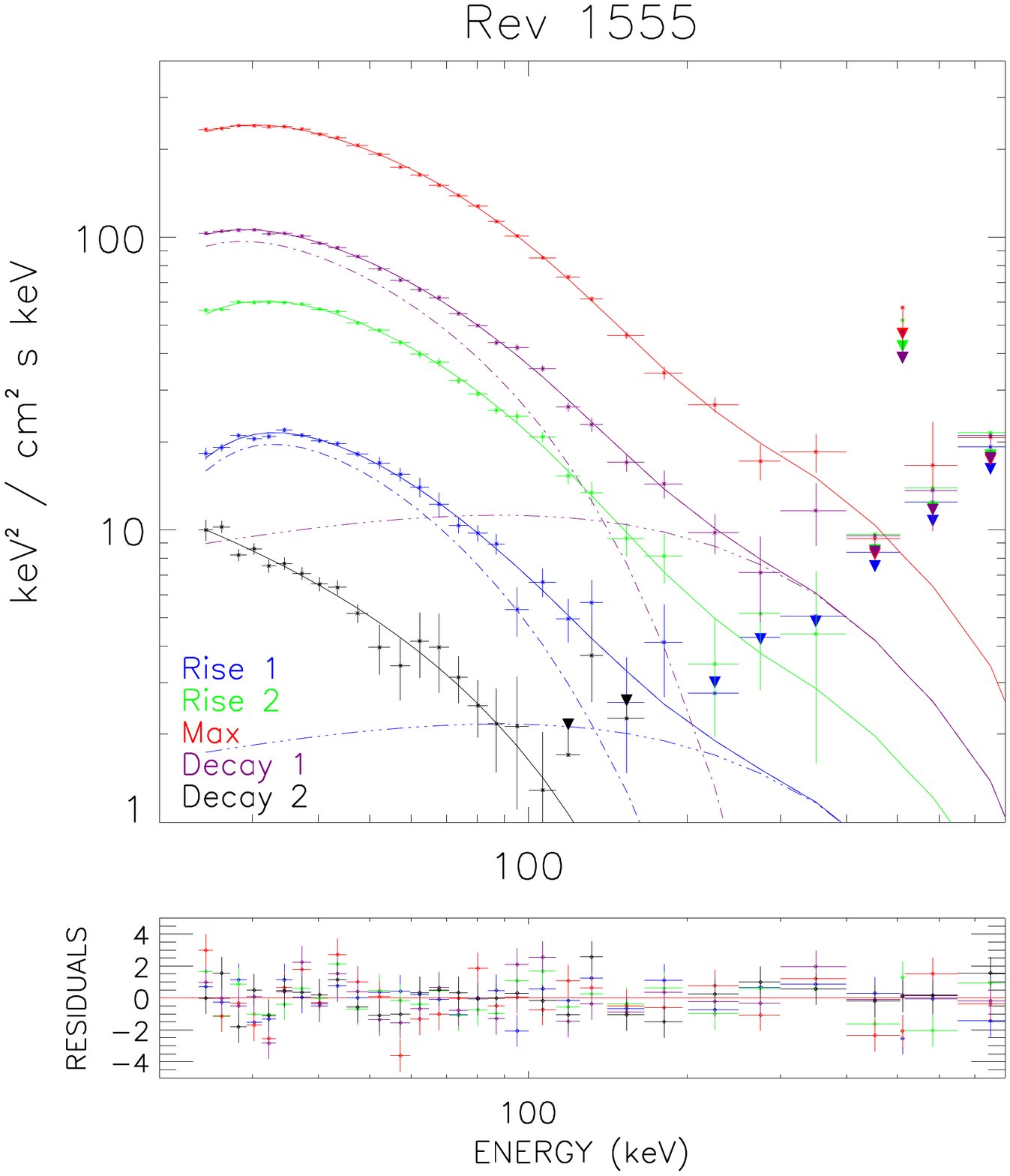}
 \includegraphics[scale=0.4]{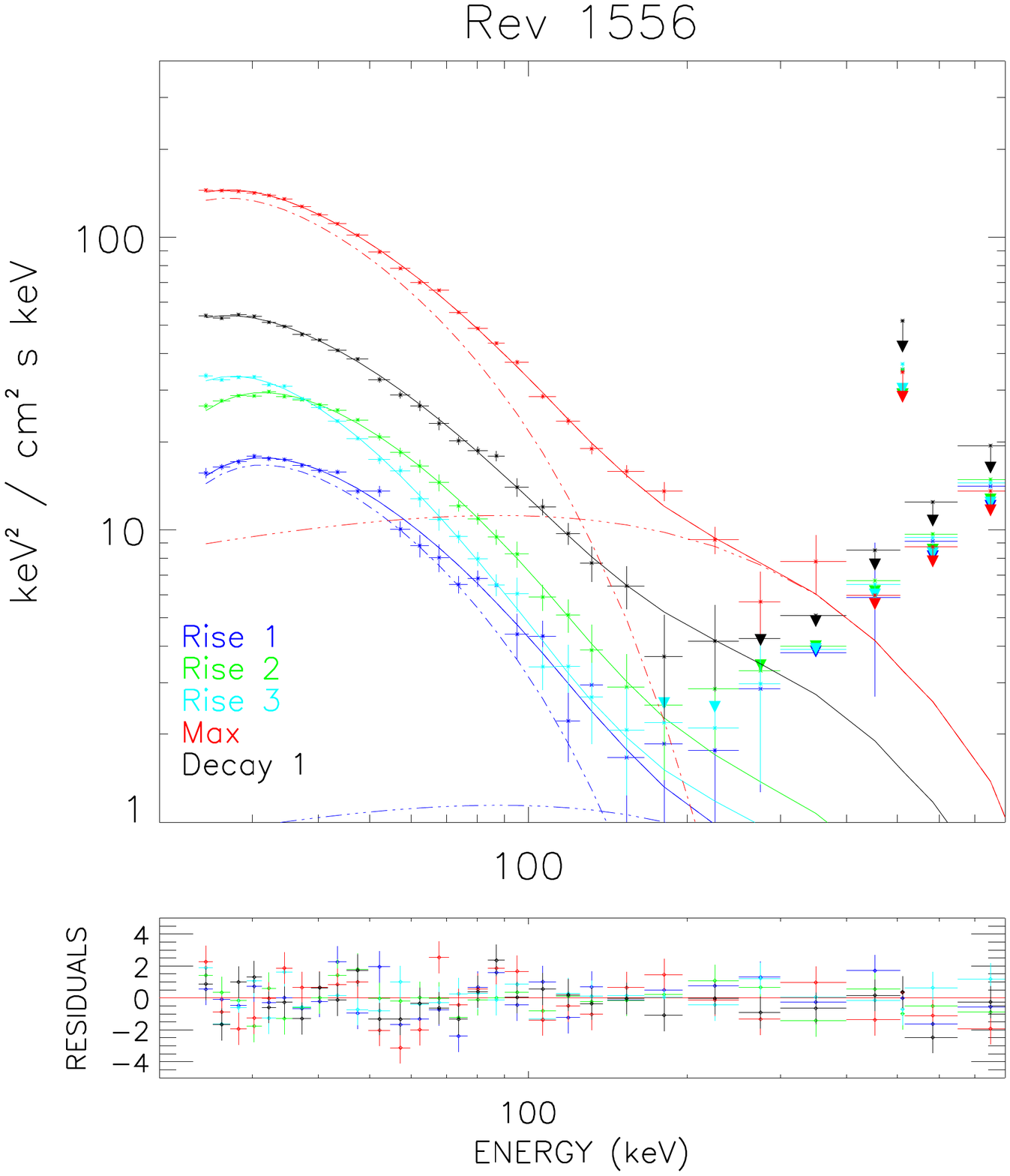} 
 \includegraphics[scale=0.4]{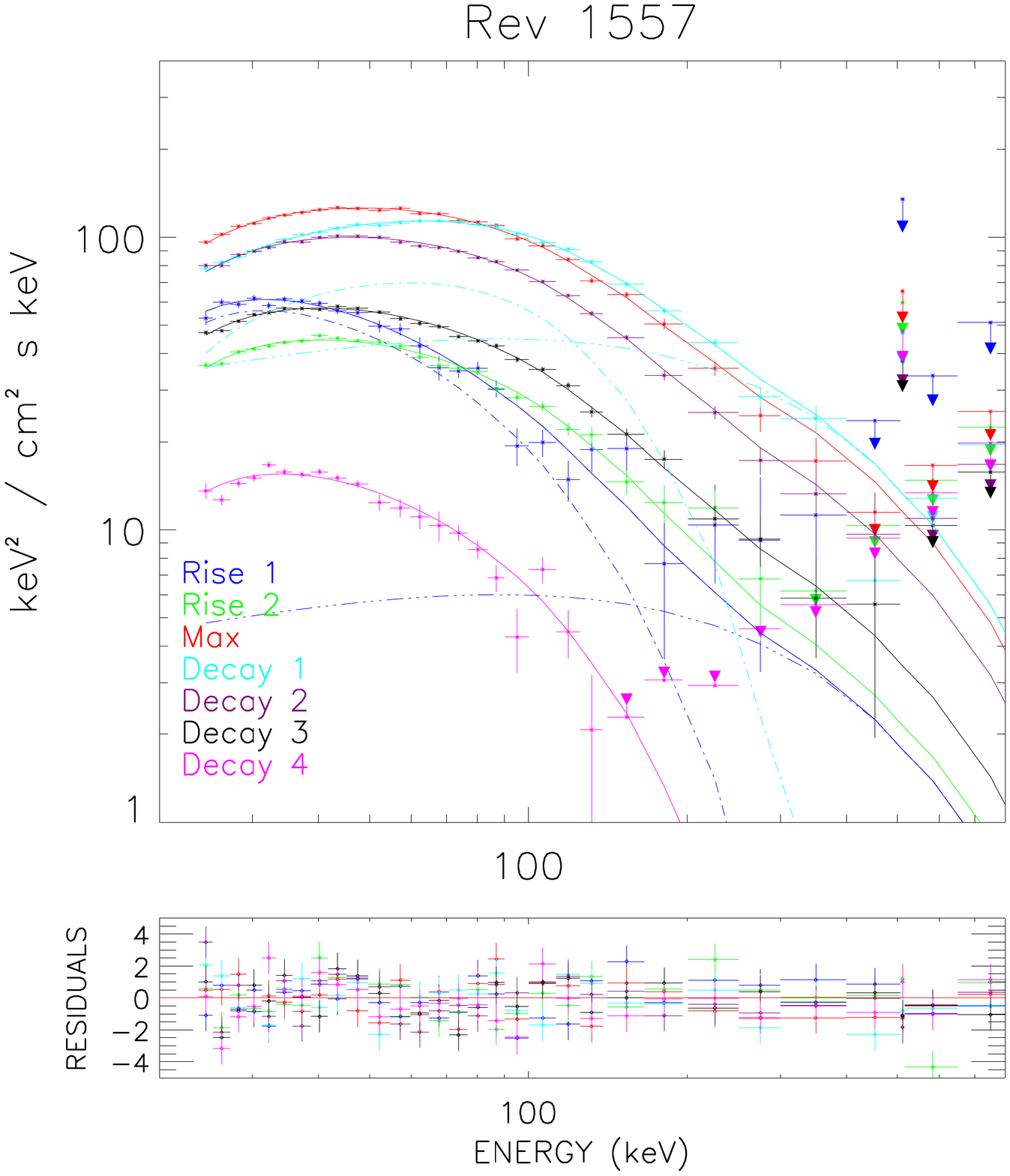}
 \caption{Spectral evolution for the 4 flares marked in fig.\ref{LCV404}.}\label{SPV404}
\end{figure}

The model considered here is, in Xspec language: wabs*compTT+cutoffpl. 
The best-fit parameters are given in Table \ref{tabV404} while the spectral evolutions during each of these flares are displayed in fig. \ref{SPV404}.
The temperature of the Comptonizing electron population can be set constant within each flare and almost constant (22 to 27 keV) from one flaring episode to the other.
Conversely, the electron density is required to vary, more or less according to the flare. As spectra are averaged over $\approx$ 1 hour, whereas the source varies on much shorter timescales, kT and $\tau$ are to be taken as indicative values of a macroscopic comportment. Moreover, these two parameters suffering from  degeneracy in the Compton fit procedure, the $\tau$ evolution might be seen more conservatively as the evolution of the Compton y-parameter.

The absorbing medium is stable ($N_h$ around 8 $\times 10^{24} cm^{-2}$) for the revolutions 1554 and 1557. For the revolutions 1555 and 1556, the absorber is thicker at the beginning of the flare (above $10^{25} cm^{-2}$), then becomes thiner. At the end of the revolution 1555 flare (corresponding to the lowest flux level among our sample), no absorption at all affects the source emission above 20 keV.

Along with the $N_h$ parameter, the source variability is driven by the optical depth and the intensity of the cutoff power law relatively to the Compton component. We define its fractional contribution, $f_{HE}$, as the ratio of the powerlaw flux over the total flux, in the 20-150 keV energy band (see last column of Table \ref{tabV404}).
The variations of these three parameters explains the spectral shape evolutions observed in fig. \ref{SPV404}. For revolutions 1555 and 1556, the high energy contribution and the optical depth values are both weak with little changes. The most impacting parameter is the absorbing column density, but it affects only the low energy part (below $\sim$ 50 keV). In contrast, for revolutions 1554 and 1557, $f_{HE}$ and $\tau$ values are higher and vary, what a more, in a correlated manner. This results in a complex variability pattern. However, the correlation observed between $f_{HE}$ and $\tau$ points out an evolution more organized than it seems.

\begin{figure}
\includegraphics[scale=0.40]{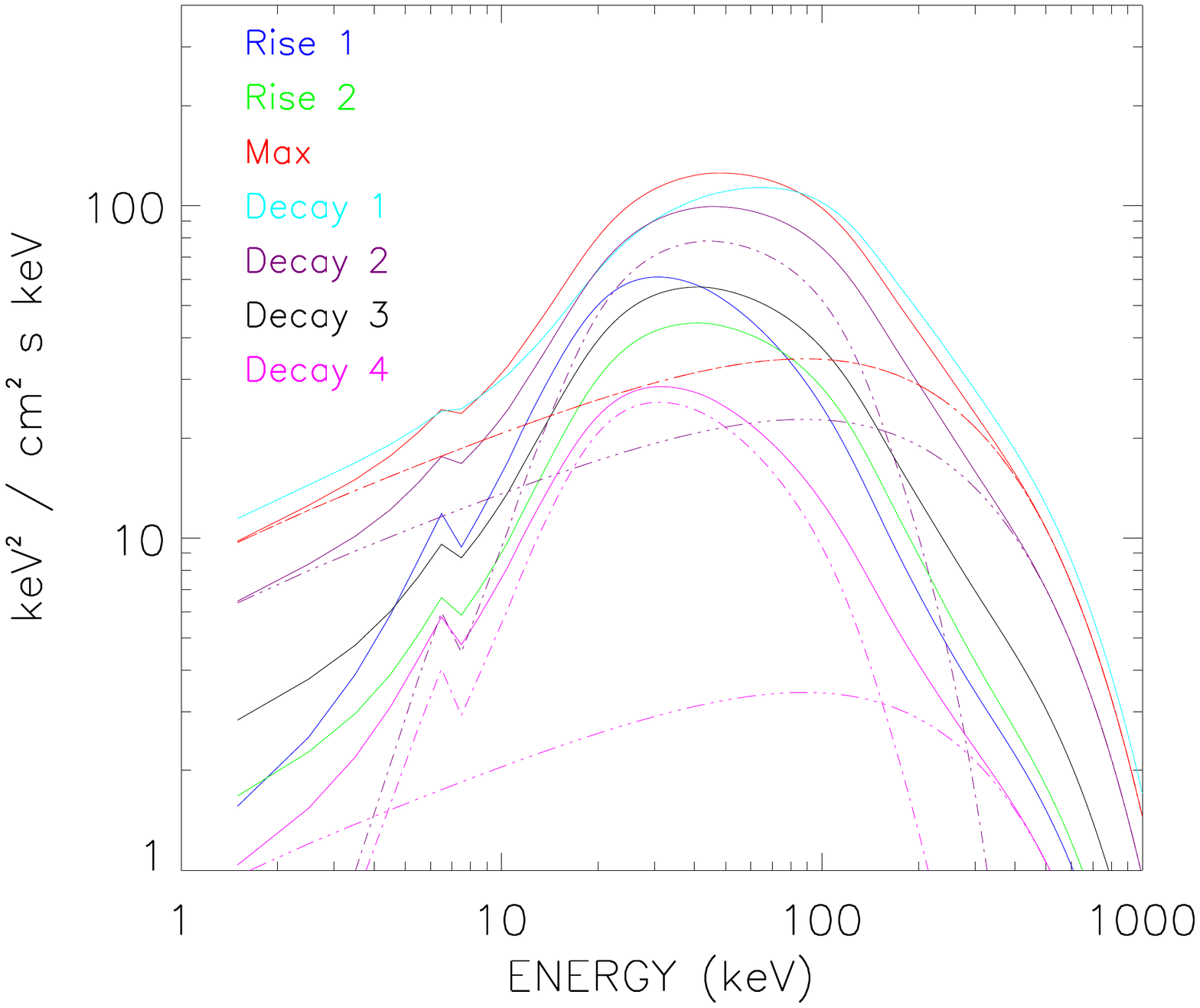}
\caption{Theoretical spectra from best fit model for revolution 1557.}\label{SPNustar}
\end{figure}

\paragraph{What about the X-ray emission?}
In the scenario explored above, the huge column density values prevent the detection of any X-ray emission related to the Comptonization region. X-ray observations are thus crucial to test the proposed model and  may help to constrain it. In particular, \textit{NuSTAR} provides invaluable information, with high quality spectra  reported  by \cite{Walton17}. There are no simultaneous observations between \textit{NuSTAR} and \textit{INTEGRAL} during flare episodes. Nonetheless, we have tried to take into account the elements deduced from \textit{NuSTAR} observations to improve our view of the GS 2023+338 geometry.
A major feature observed in the X-ray band is the reflection component and more particularly the 6.4 keV iron line. We thus consider an alternative model with an important reflection component and no or thin absorbing medium. We use a scenario based on a pure reflected emission to account for the emission between 20 and 100-150 keV, with still a cutoff power law for the higher energy emission (Xspec command: model=reflect*compTT+cutoffpl, with rel\_refl parameter fixed to -1). It stands e.g. for a source geometry where the Comptonizing region is seen by reflection on the inner part of an optically thick torus, while the second component (for instance produced by a jet) is emitted from outside the torus.  This model provides an acceptable description of the SPI data, even if $\chi^2$ values are slightly higher than for the first scenario. Despite the absence of simultaneous data, we have used  the Xspec tool to plot the model spectra in the \textit{NuSTAR} band, in order to compare qualitatively the X-ray emission expected from our results and the \textit{NuSTAR} observations.
Fig.\ref{SPNustar} displays the theoretical spectra expected from the \textit{INTEGRAL} best-fit parameters obtained from the spectral fitting of the first flare of the revolution 1557. A comparison with spectra presented in \cite{Walton17} indicate that such a model could reconcile X-ray and hard X-ray observations, at least quantitatively. In particular,
the fluxes predicted in the 3-79 keV energy band (between 10 and 35 $10^{-8}~erg~cm^{-2}~s^{-1}$) are quite commensurable  with those reported in \cite{Walton17}.

To conclude this study, the hard X-ray spectra observed from GS2023+338 require a peculiar configuration of the emitting region. A first hypothesis relies on a Compton component with a very thick medium around the source and thus imposes that the observed X-ray  photons are emitted from a separate region. A second scenario corresponds to a geometry where the primary Compton emission is hidden: we observe a pure reflected component. Both models are completed witha second component to describe the high-energy tail. Various solutions in-between these two extreme scenarios (with the presence of both reflection and absorption) might provide similarly acceptable description of the data.  The main point to note is that, in all cases, the primary emission below 10 keV is closely related to the high-energy tail observed above 200 keV. Indeed, \cite{Walton17} propose that the  X-ray continuum  observed during the flares (modeled with a cut off power law) comes from transient jet ejection events. Its identification with the component emerging above 200 keV, at least during the flaring periods of GS2023+338, deserves a further building.  Yet, such a scenario would significantly change our view of the mechanisms at work in the low and high-energy photon production at least during this intense and extraordinary outburst.

\subsection{MAXI J1820+070}

\begin{figure}
\includegraphics[scale=0.5]{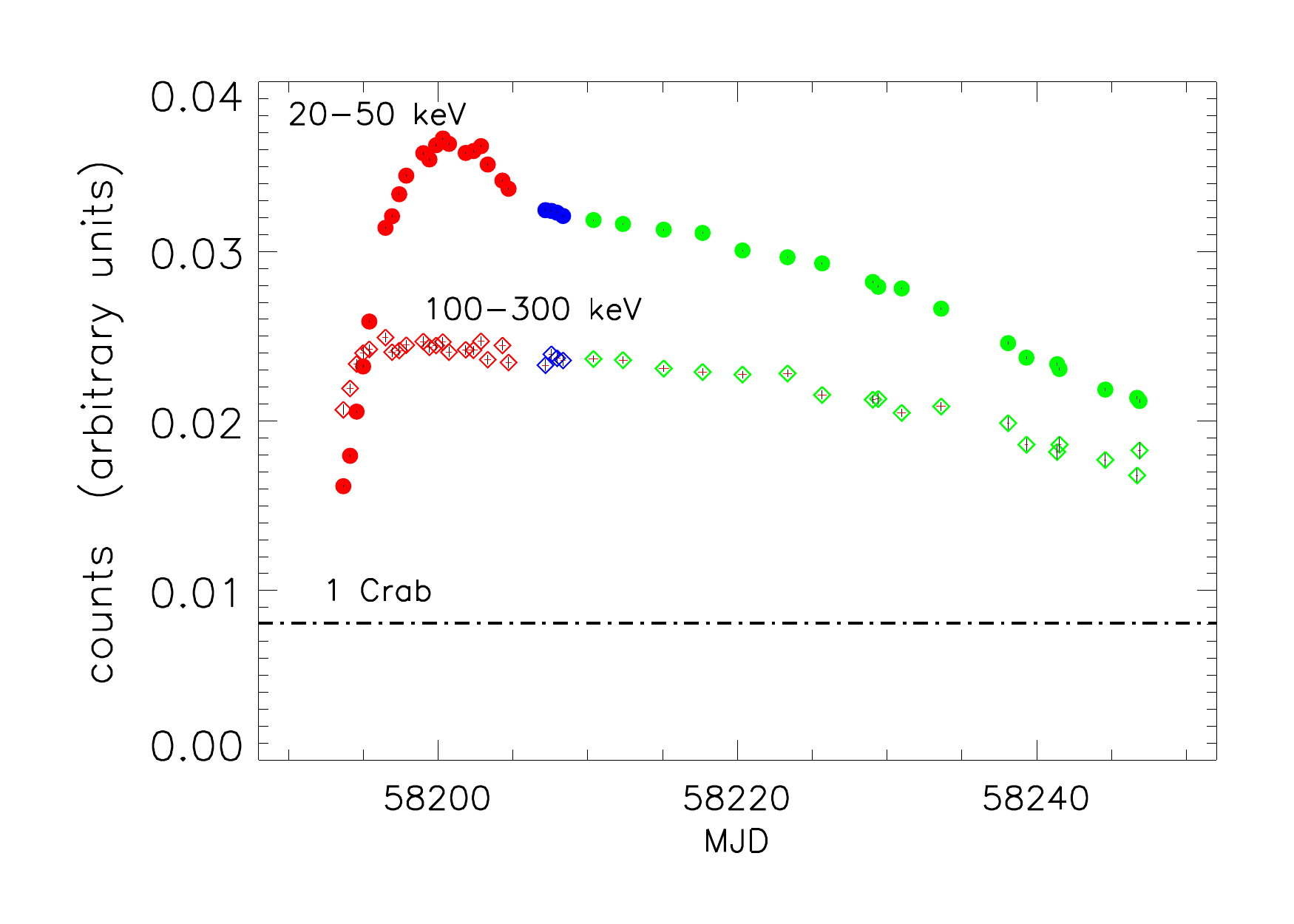}
\caption{Light curves of MAXIJ1820+070 in two energy bands. The dot-dashed line corresponds to 1 Crab.}\label{MaxiLC}
\end{figure}

MAXI J1820+070 has been discovered by the MAXI/GSC instrument on the 2018, March 11 \citep{Maxi}. The \textit{INTEGRAL} mission has started a dedicated ToO campaign on the 16th of March. The log of the observations is presented in Table \ref{MAXIobs}.
\paragraph{Light curves and Hardness ratios}
 Fig. \ref{MaxiLC} displays the temporal evolution of the source in two energy bands (20-50 keV and 100-300 keV) for the \textit{INTEGRAL} follow-up period (2018-03-16 to 2018-05-08 or MJD 58193 to 58246).
 
\begin{figure}
\includegraphics[scale=0.50]{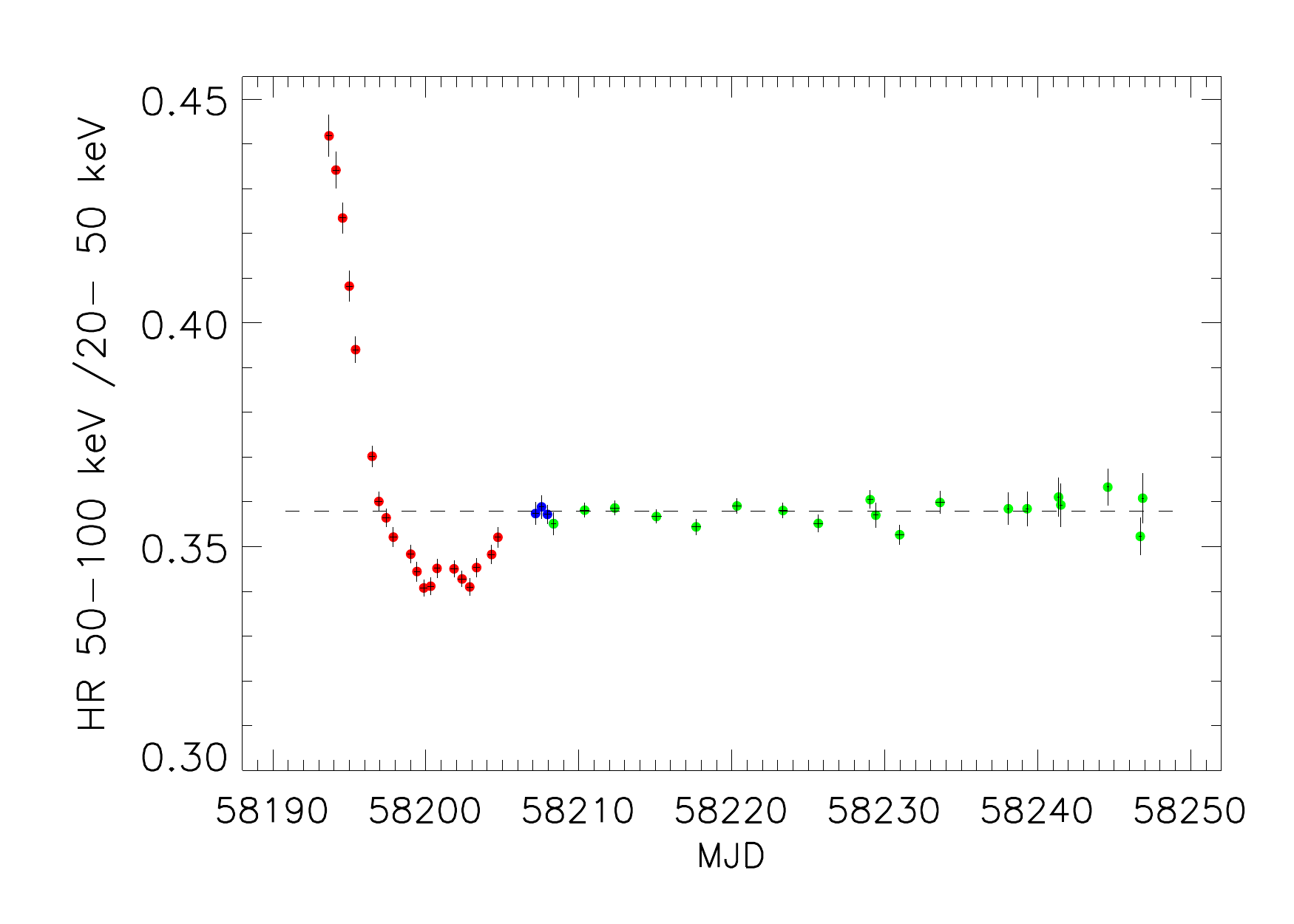}
\includegraphics[scale=0.50]{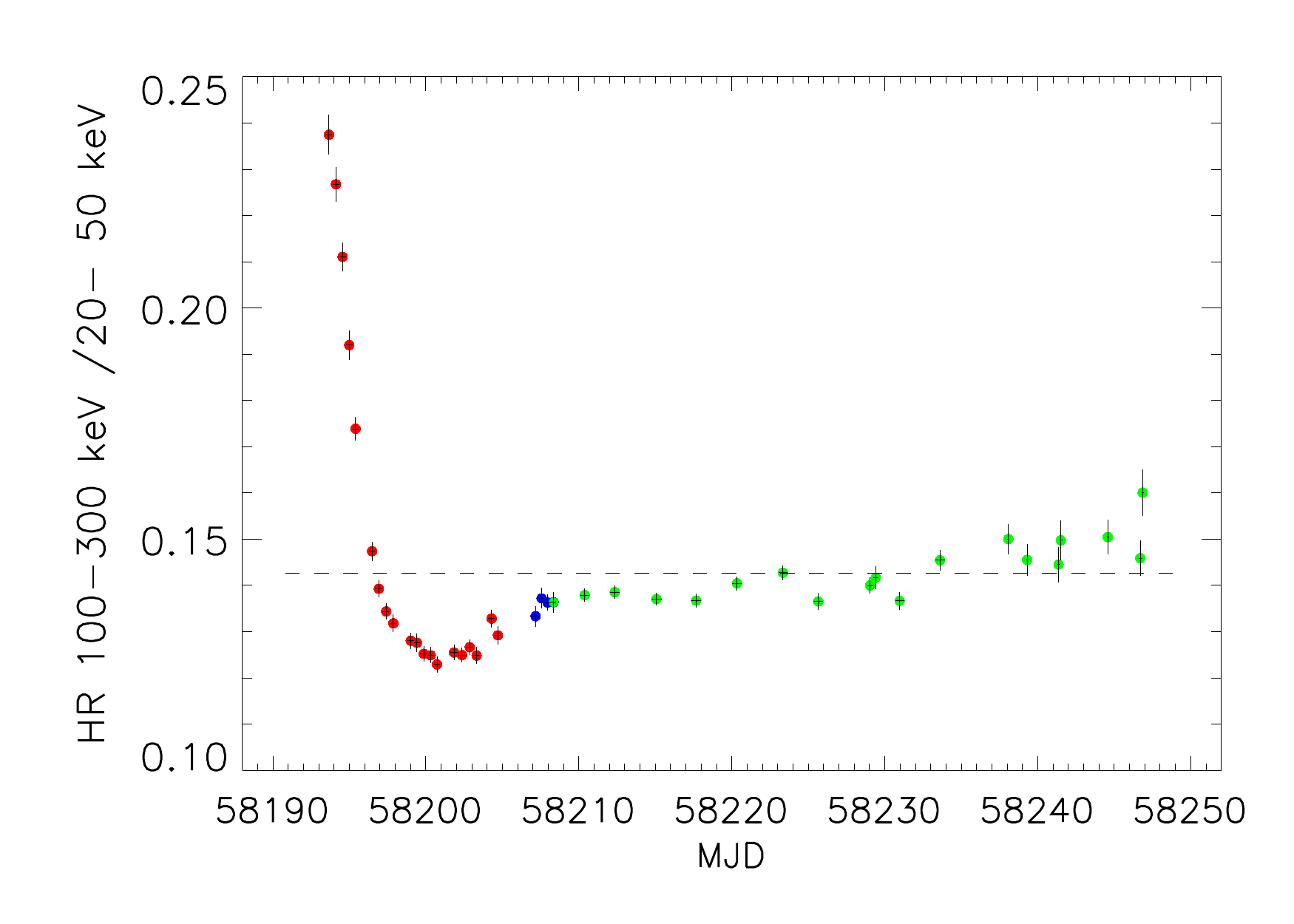}
\caption{Hardness ratio evolution of MAXIJ1820+070}\label{MaxiHR1}
\end{figure}
\begin{figure}
\includegraphics[scale=0.50]{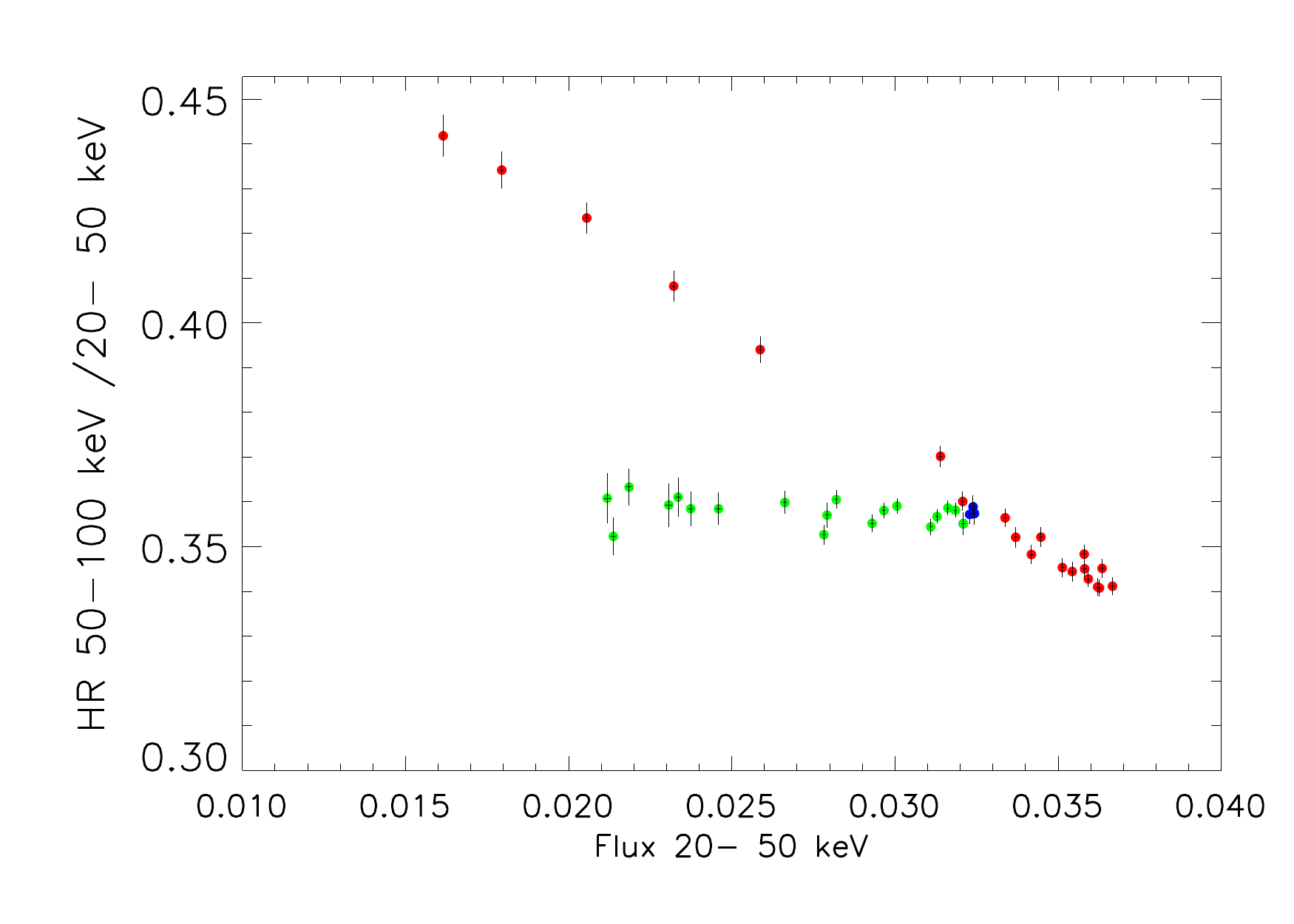}
\includegraphics[scale=0.50]{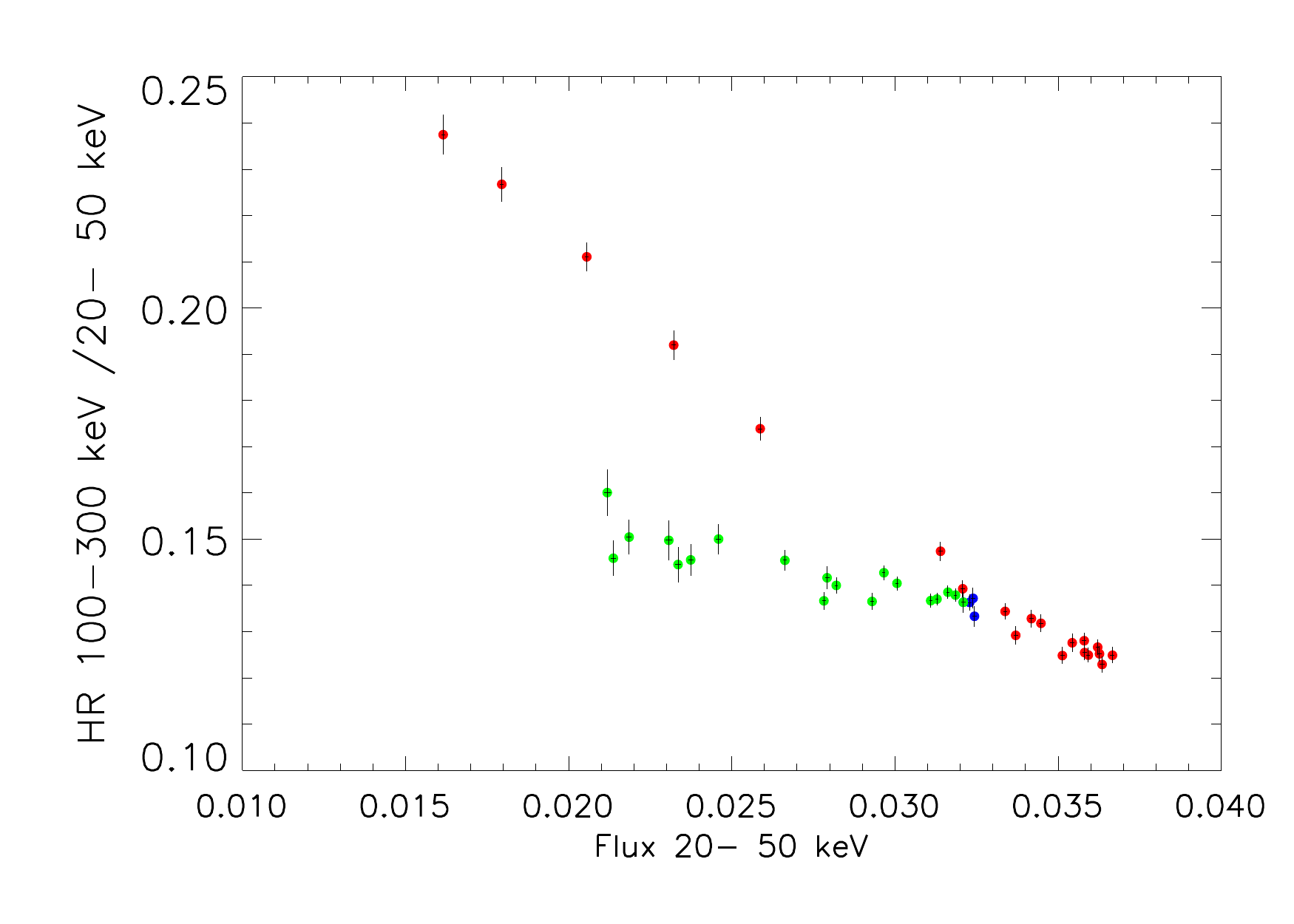}
\caption{Hardness ratio versus flux evolution of MAXIJ1820+070.}\label{MaxiHID}
\end{figure}

  
The source intensity presents a smooth rising, reaching its maximum at $\approx$ 4 Crab, in the 20-50 keV energy band, $\approx$ 12 days after its first detection. Then, its intensity decreases regularly with a marked slowing down in the decay around MJD 58206.
The high-energy emission (100-300 keV) reaches its maximum before the low energy (MJD 58195 and 58200 respectively). Moreover, during the bump displayed by the low energy emission light curve, the high-energy emission remains almost stable and then decreases slower than the low energy part.
This decoupling between the low and high-energy parts is illustrated in fig. \ref{MaxiHR1} with the evolution of the source hardness versus time.
Another way to see the spectral evolution is proposed in fig. \ref{MaxiHID} through HR-flux diagrams.

The source emission clearly behaves differently before and after MJD 58206. On each figure, we plot in blue color the points just after this date. Before MJD 58206, the spectral shape at low energy (20-100 keV) softens when the flux increases and conversely, with a very close relation between the flux and the hardness. After MJD 58206, the spectral shape below 100 keV remains unchanged (HR = constant, left panels in fig. \ref{MaxiHR1} and \ref{MaxiHID}).
At higher energy, the anti-correlation between the hardness and the flux is observed all along the burst (right panel of fig. \ref{MaxiHID}), but the evolution follows a separate track after MJD58206.

\paragraph{Spectral analysis}
\begin{figure}
\includegraphics[scale=0.4]{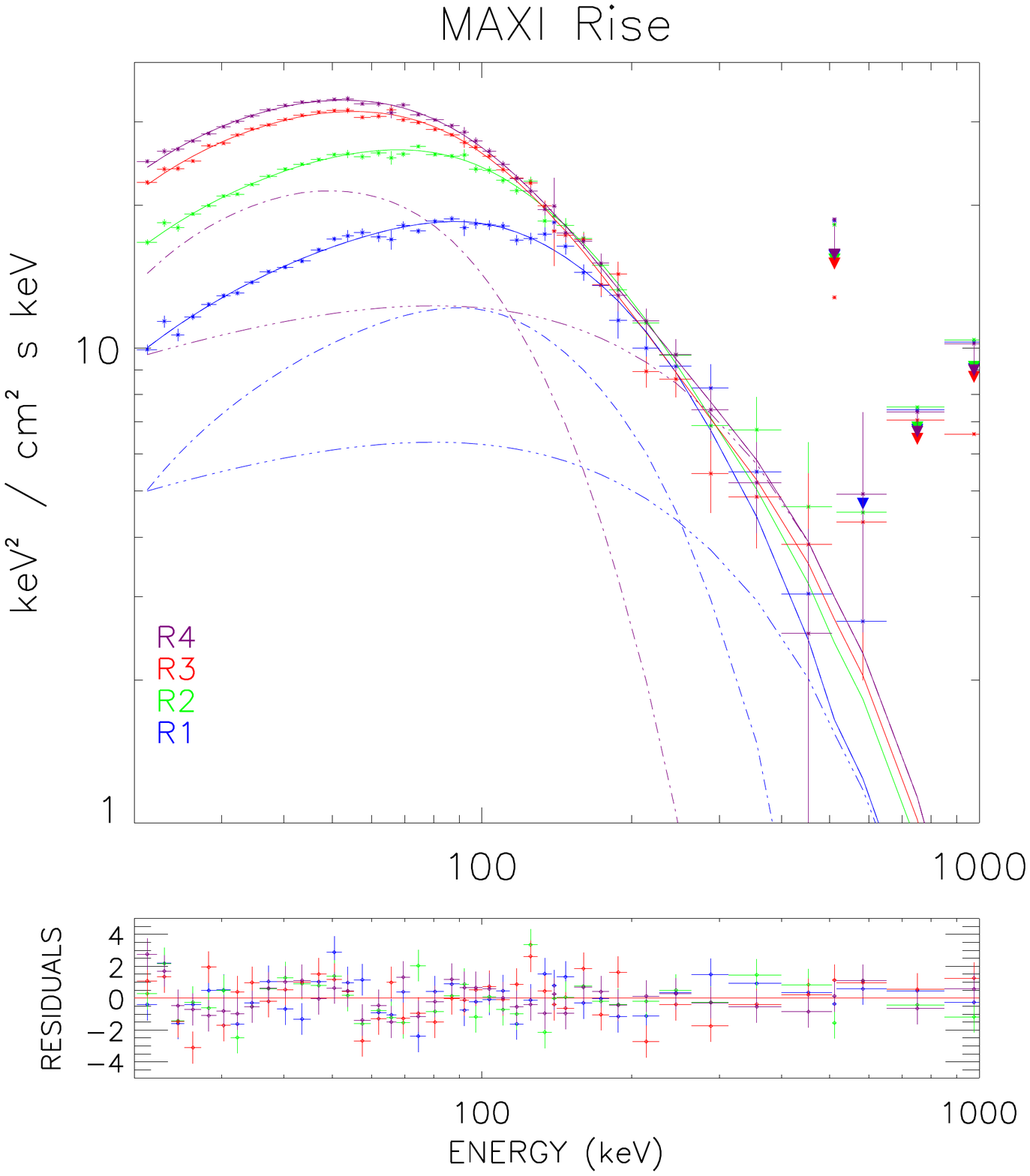}
\includegraphics[scale=0.4]{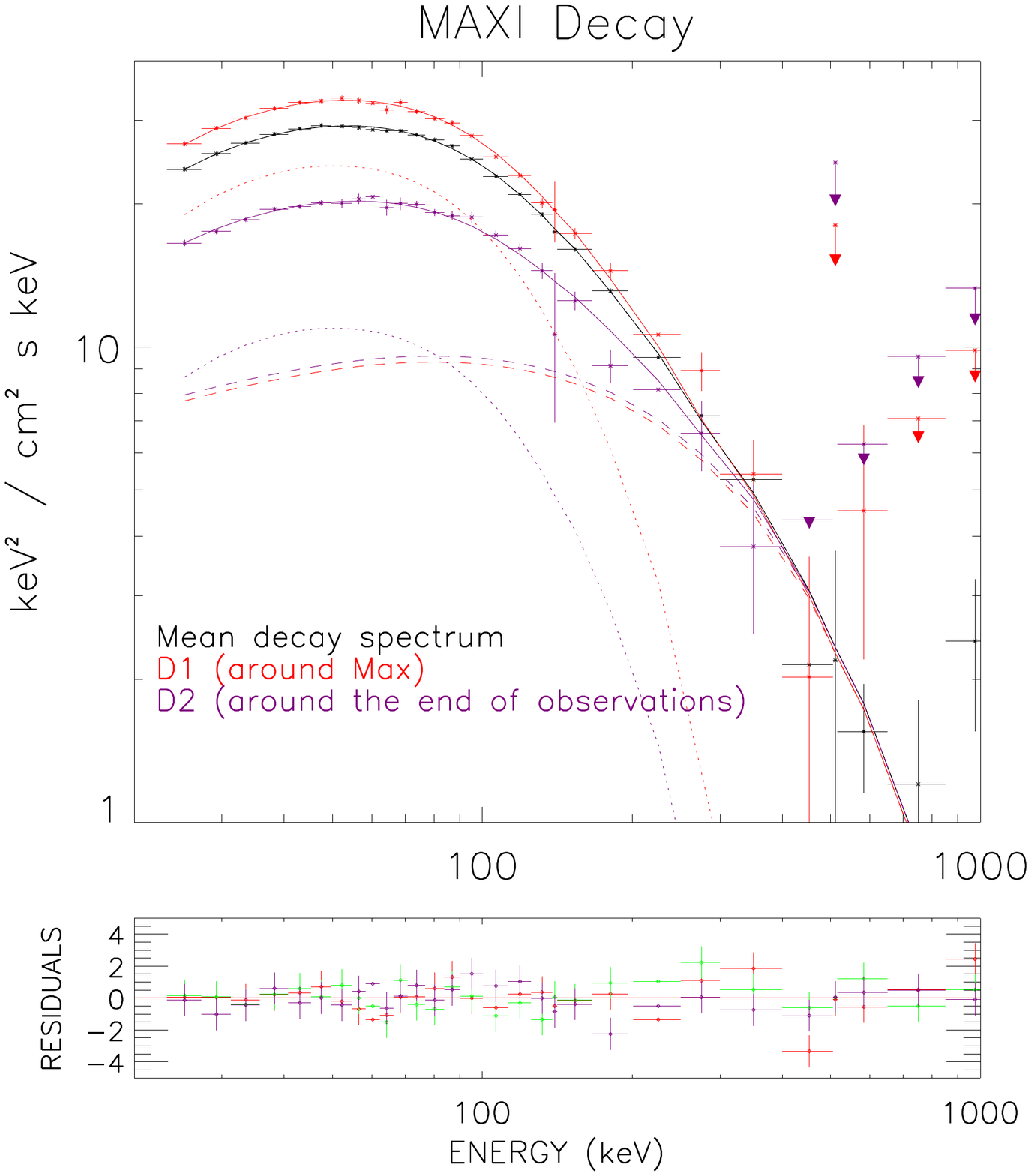}
\caption{Spectral evolution of MAXIJ1820+070.}\label{MaxiSP}
\end{figure}
\begin{figure}
\includegraphics[scale=0.5]{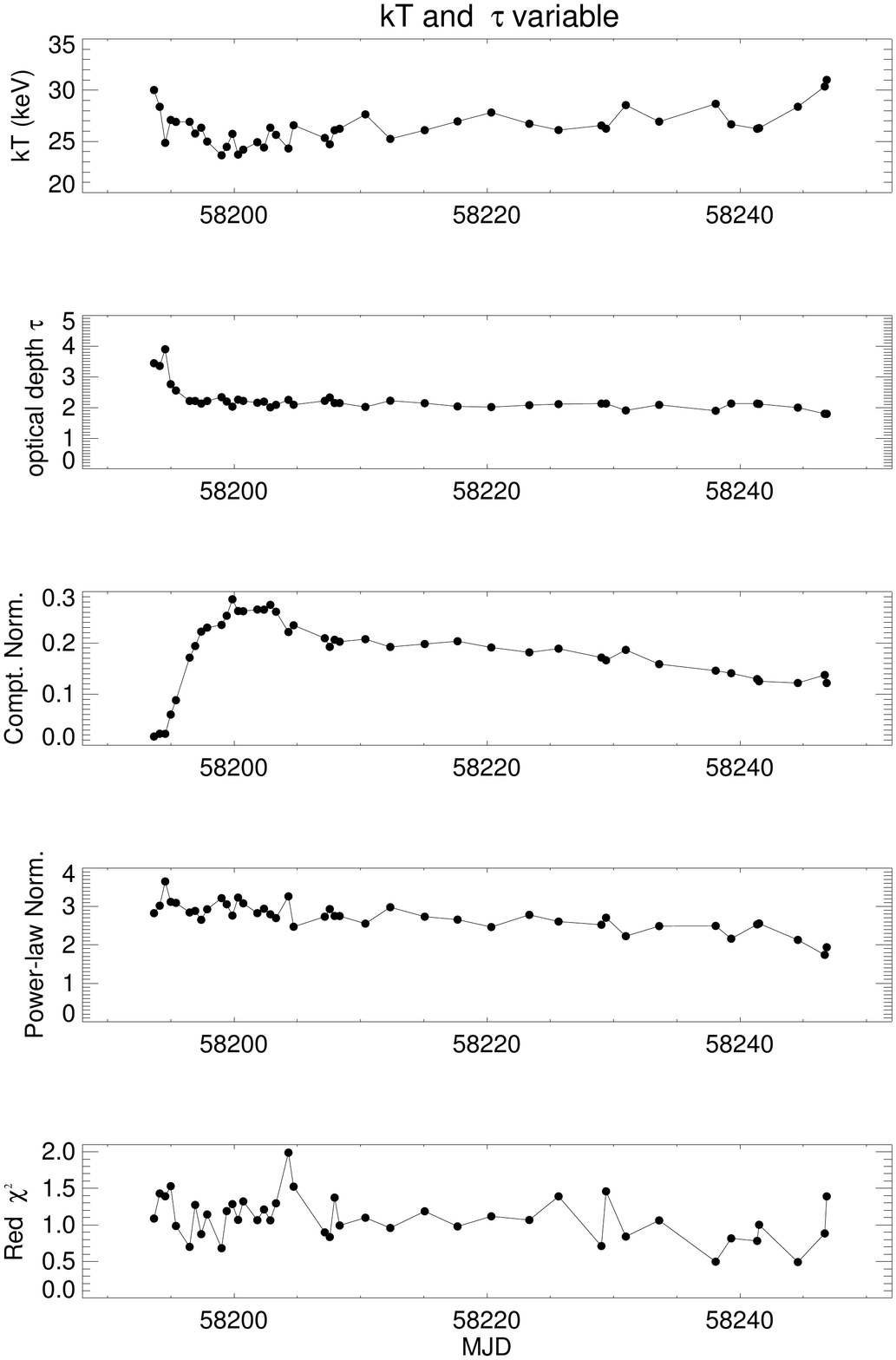}
\caption{Evolution of spectral fit parameters MAXIJ1820+070, reflection factor fixed to 1.}\label{Maxiref1}
\end{figure}
\begin{figure}
\includegraphics[scale=0.40]{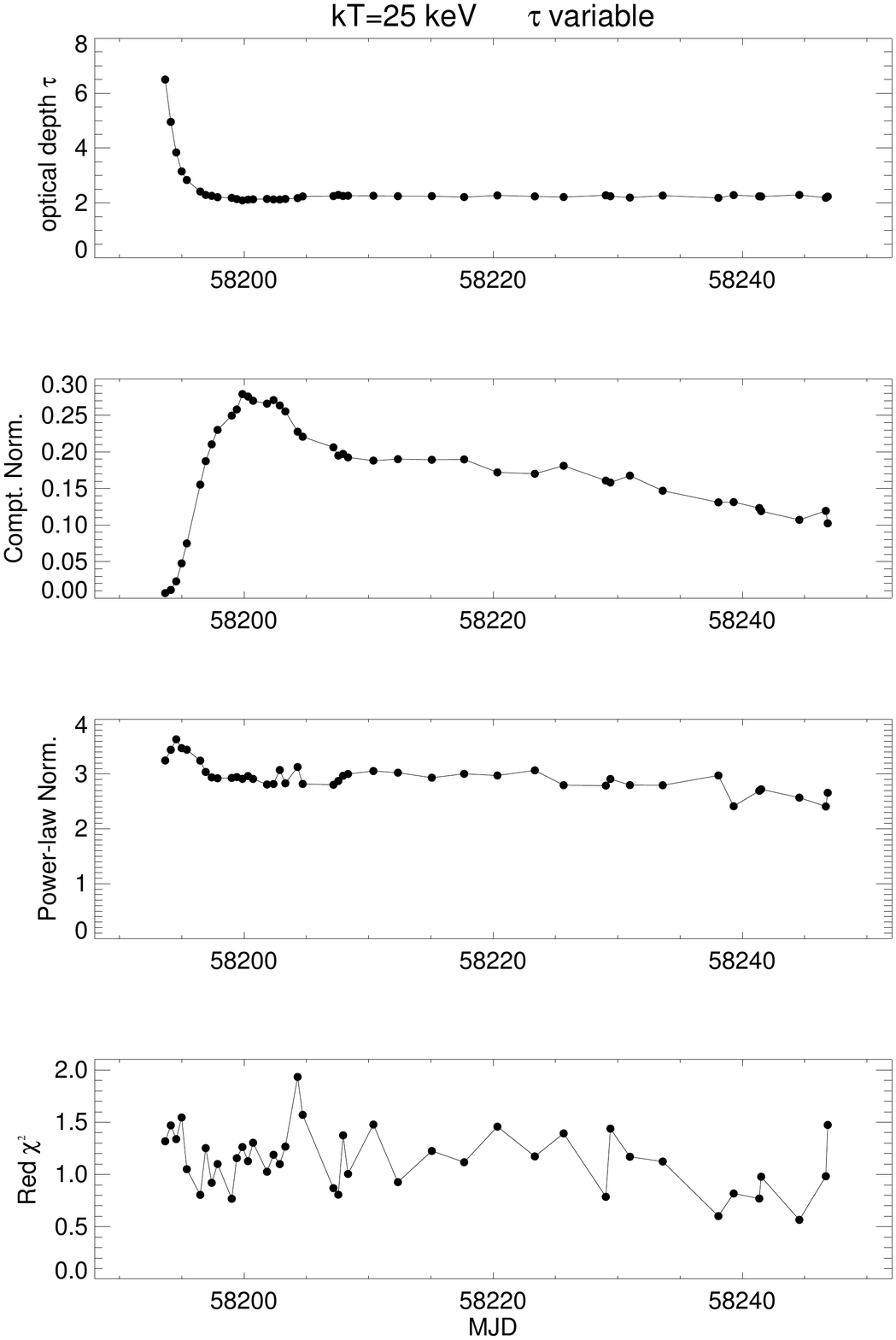}
\includegraphics[scale=0.40]{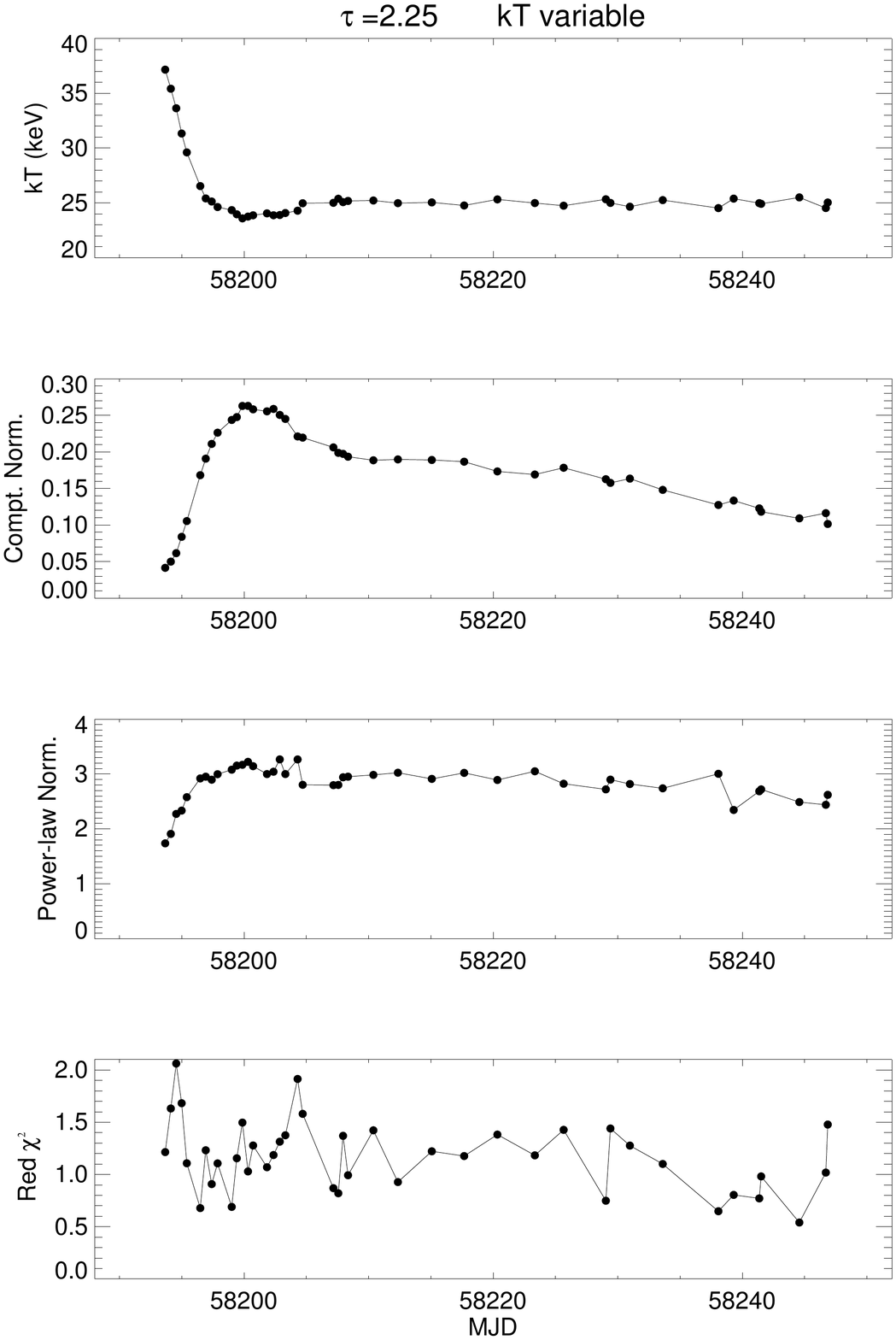}
\caption{Evolution of spectral fit parameters MAXIJ1820+070, left panel: kT fixed, right panel: $\tau$ fixed.}\label{Maxirefkt}
\end{figure}
To explore more precisely the evolution of the source emission, we have built the spectra corresponding to each of the 42 temporal bins used in fig. \ref{MaxiLC}. A few of them are displayed in fig. \ref{MaxiSP} to illustrate the spectral behavior of the source along its outburst. It makes clear the dissimilarity of the  evolution of the source emission during the rising (left panel) and decay (right panel) phases.  At the beginning of the outburst, the peak energy shifts from 100 to 50 keV. It remains at this value during the decay. \\
All the above results suggest that two different components contribute to the observed emission.
First tests on a sample of spectra confirm that a Comptonization law fails to describe the high-energy data, beyond the Comptonization roll-off, i.e. above 100-150 keV. We thus performed a global spectral analysis based on a two component model: a Comptonization law, with a reflection component, and a fixed shape cutoff powerlaw (photon index = 1.6 and energy cutoff = 200 keV). In this model, the free parameters are: the reflection factor (Rref), the temperature and optical depth of the Comptonizing population (kT, $\tau$) and the normalizations of both components.
It is not possible to constrain together the three shape parameters. Fixing the reflection amplitude to 1 appears as a reasonable assumption.
In this framework, we fit the 42 source spectra to obtain the evolution of kT and $\tau$ along the outburst. The findings are presented on fig. \ref{Maxiref1} together with the resulting $\chi^2$. After a first period where the source parameters evolve dramatically, both kT and $\tau$ display a more stable comportment, the first one slightly increasing, the latter slightly decreasing. This anti-correlated evolution reminds the kT/$\tau$ degeneracy, leading us to fix still one more parameter. The results corresponding to both solutions (kT fixed to 25 keV, left panel and $\tau$ fixed to 2.25, right panel) are presented in fig. \ref{Maxirefkt}.

The $\chi^2$ values remain comparable to those obtained with both parameters varying, except for the very first observations when $\tau$ is fixed. Indeed, the source evolution during \added{the first} revolution \deleted{1931} is more pronounced than during the remaining of the outburst. In our fits, it results in large values of kT or $\tau$ (or Rref, if we fix kT and $\tau$), suggesting that a more complex scenario has to be considered for this period.
Looking now at the powerlaw component contribution (2nd panel from bottom), we notice that the best fit normalization is not significantly affected by the low energy model, except again during the first revolution. 
Indeed, the powerlaw amplitude evolution basically corresponds to that of the 100-300 keV flux displayed in fig. \ref{MaxiLC}, with a moderate increase and limited decrease.

A number of sources present such a hard tail, independent of the Comptonization emission. The best example is Cyg X-1 with the detection of a strong polarization, \citep{PL2011,CygpolarEJ}, strongly suggesting that the high-energy component is related to the presence of a jet. For MAXI J1820+070, another argument  supports this hypothesis  since a significant radio emission has been reported  during the period covered by our observations. Moreover, this emission does not vary too much, in contrast to the X-ray emission. It declines by less than 15\% from MJD58195 to MJD 58225 \citep{Radio}, similarly to the 100-300 keV flux reported here. MAXI J1820+070 is thus clearly another example, where a common population for radio and high-energy emission should be considered.

\section{Conclusions}
In the first part of the paper, we discuss the generation of spurious events by the analog front-end electronics. We establish that a tiny fraction of low energy photons is displaced towards high energies,  during the recovery time of the electronics after a saturating event. Then, we demonstrate that, thanks to the capabilities of the SPI triggering scheme, the phenomenon occurs only in the SE flagged events while the PE flagged events are immune to it.
We show that the presence of these spurious events has a dramatic effect on the spectra of celestial sources above 400 keV, when these sources exhibit a high, low-energy flux level (above 1 Crab), and that simple rules concerning SPI event selection lead to systematic error free spectra.\\
In a second part, we study the spectra of three bright hard X-ray sources using these simple recommendations.
The Crab Nebula spectral shape evolves smoothly from a slope of $\approx 2$ in the low energy range (20-50 keV) towards a slope of 2.2-2.3 above 150 keV. The best description of the data is obtained when using the Band model \citep{band93}, and the spectral shape is remarkably stable over 16 years of observations.\\
We then analyze four flares occurring during the 2015 outburst of GS2023+338 (=V404 Cygni). While the spectral evolution appears complex and varies from one flare to the other, the emission can be described by two components with a limited number of free parameters. Moreover, our results suggest that the hard tail dominating the source emission above
150-200 keV may be the same component that the primary continuum observed in the X-ray range.\\
The last study is devoted to another transient X-ray binary, discovered in 2018 March, MAXI J1820+070.
As for GS2023+338 and other X-ray binaries, two independent components contribute to the spectral emission of MAXI J1820+070 in the hard X-ray domain. An interesting result is the decoupling of the temporal evolution of these two components.  Also, the source behavior changes a few days after the maximum of the X-ray emission. While the spectral shape evolves dramatically during the rising phase, it remains very stable (kT and $\tau$ constant) all along the decay, when the source intensity slowly decreases.

\section*{Acknowledgments} The \textit{INTEGRAL} SPI project has been completed under the responsibility and leadership of CNES.  We are grateful to ASI, CEA, CNES, DLR, ESA, INTA, NASA and OSTC for support.

\appendix
\section{More on SE versus PE: the PE/(SE+PE) curve}
\begin{figure}
\includegraphics[scale=0.50]{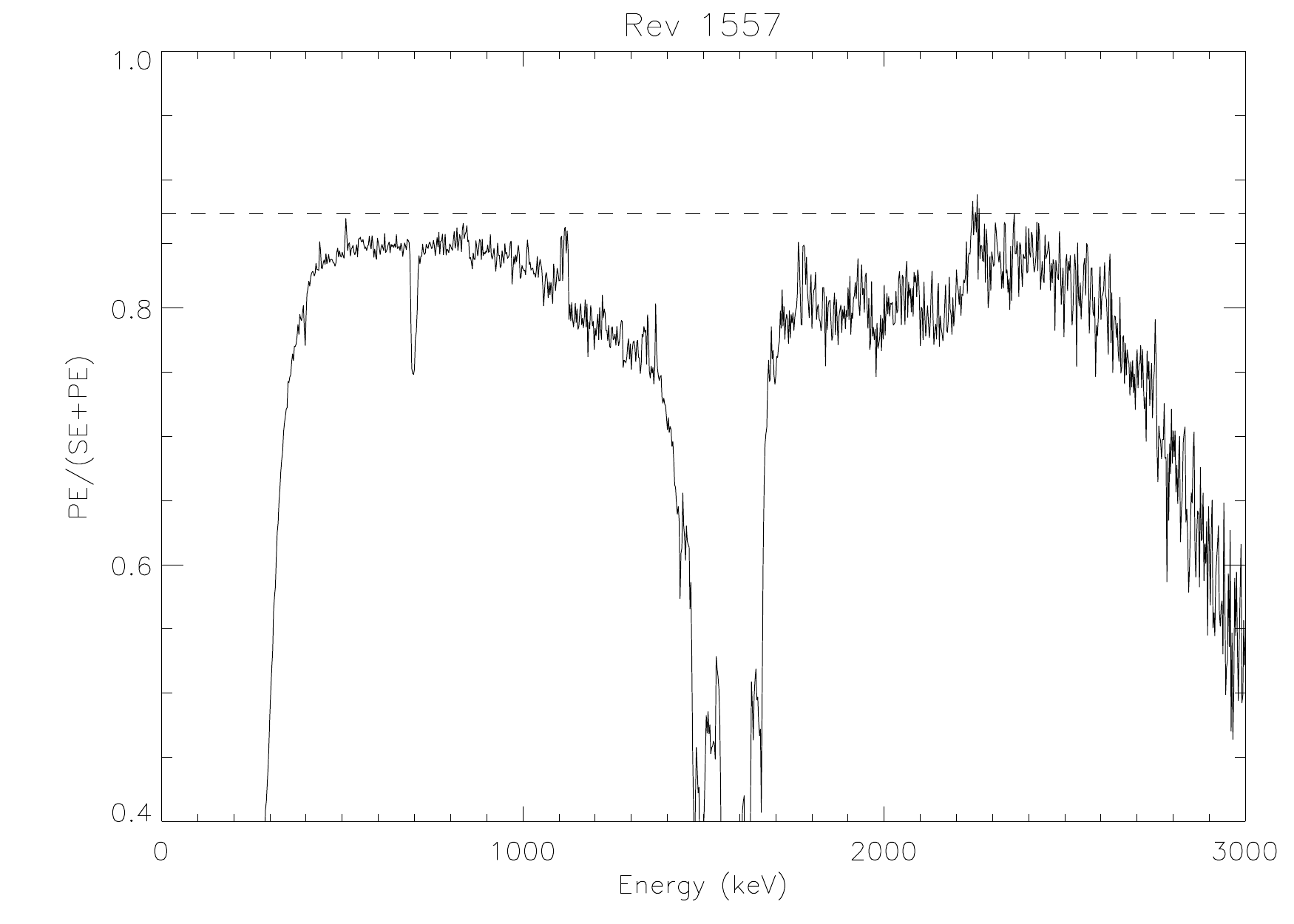}
\caption{Ratio PE/(SE+PE) versus energy. The dashed line corresponds to the actual PSD efficiency.}\label{PESEratio}
\end{figure}

In order to investigate deeper the PE versus SE spectra, the ratio PE/(PE+SE) versus energy, where PE and SE represent count rates of PE events and  SE events respectively, is plotted in figure \ref{PESEratio} for revolution 1557. It appears that this ratio is below the PSD efficiency value measured from background lines. A smaller value comes from a larger denominator. It thus implies that  the SE spectrum contains too many events, that are "not confirmed" by the PSD electronics, as explained in section 2. In particular, we can  identify the spectral region around 1000-1600 keV (corresponding to the enhancement already mentioned in figure 1), that contains a large fraction of  events not seen by the PSD device (known  as 'electronic noise'), leading to very low values of the computed ratio, whereas the PSD efficiency measured with physical lines remains at the 87\% level. Another interesting feature concerns the   693-705 keV area: the broad shape of the 691 keV line from ${}^{72m}\text{Ge}$ is due to an inelastic neutron scattering and a delayed emission ($\tau$=444 ns). The neutron scattering, the delayed emission and the speed (40 ns) of the PSD trigger explain that PSD electronics miss a large fraction of  the photons corresponding to this "line".
The ratio PE/(PE+SE) is also reduced (relatively to the efficiency measured  in physical lines) by the fraction of  double events, missed due to the dead time, and thus classified as SE (see the section devoted to ME events).
 It is beyond the scope of this paper to investigate all the features of this PE/(SE+PE) curve, but we state that this cannot be used to derive precisely the efficiency of the PSD electronics. The latter should be derived using the photo-peak from background lines corresponding to the decay of elements producing a single photon, in order to avoid multiple detector interactions.

\section{Recommendations on event selection for a reliable analysis}

Following the study presented in section 2, we know that the SPI single event datasets contain "spurious" events while PE (PSD flagged) events do not. Hence, when it is possible i.e. in the PSD electronics energy domain, one must remove SE  events  to get  reliable results. Thus, we recommend:
- To use PE events only for  energies between  the PSD thresholds (LLD and ULD).\\
Note that the PSD thresholds have changed during the mission (see table \ref{PSDth}).\\
- To apply a correction factor of the order of 15\% (see table \ref{tabPSD})  to take into account the dead-time of the PSD electronics.\\
- To build the background data with the same event selection criterion than the analyzed  dataset. Moreover, the PSD threshold configuration should be the same for background and source observations.\\
- When the PSD low energy threshold is above 400 keV, and for strong sources (above 1 Crab): to compute an empirical correction factor, Fcorr, corresponding to 0.15\% of the  source flux at 60 keV. To compare Fcorr to the statistical error of the source fluxes below the PSD LLD. If both values are commensurable, it will be wise to subtract Fcorr to flux values obtained below the PSD LLD, to take into account the contribution of spurious events, even if this is only an approximation.\\
A last point concerns SE event reliability above the high-energy threshold of the PSD electronics ($\approx 2.7$ MeV).
First, one has to mention that SPI analog electronics switches between two gain values  according to the event energy, the threshold being around 2 MeV. Thus, the behavior of the electronics, in reaction to saturating events, might be different for energies above 2 MeV from what we observe for energies below 2 MeV. Unfortunately, the source fluxes above 2 MeV is too low to perform the appropriate study and another approach is needed.

\begin{deluxetable}{ccc}
\tablecaption{PSD thresholds \label{PSDth}}
\tablehead{
\colhead{Revolutions} &\colhead{LLD (keV)} &\colhead {ULD (keV)}
}
\startdata
22-47 & $\approx400$  & $\approx2200$ \\
48-419 & $\approx650$ & $\approx2200$ \\
420-445 & $\approx750$ & $\approx2200$ \\
451-975 & $\approx600$ & $\approx2200$ \\
982-1169 & $\approx650$ & $\approx2700$ \\
1170--now & $\approx400$ & $\approx2700$ \\
\enddata
\end{deluxetable}

\begin{deluxetable}{ccccccccccc}
\tabletypesize{\scriptsize}
\tablecaption{PSD efficiency measured using background lines for a few \textit{INTEGRAL} revolutions\label{tabPSD}}
\tablehead{
\colhead{Revnum / E} & \colhead{271keV} & \colhead{310keV} & \colhead{352keV} & \colhead{403keV} & \colhead{438keV} & \colhead{584keV} & \colhead{1117keV} & \colhead{1764keV}
& \colhead{2223keV} & \colhead{2754keV}
}
\startdata
43 & 0.407 & 0.672 & 0.80 & 0.852 & 0.874 & 0.881 & 0.882 & 0.874 & 0.839 & 0.43\\
1557 & 0.275 & 0.555 & 0.763 & 0.841 & 0.859 & 0.869 & 0.872 & 0.874 & 0.8311 & 0.713 \\
1856 & 0.261 & 0.538 & 0.757 & 0.852 & 0.837 & 0.847 & 0.853 & 0.849 & 0.815 & 0.707 \\
1933 & 0.25  & 0.528 & 0.761 & 0.81  & 0.83  & 0.845 & 0.845 &  0.836&  0.779  & 0.693 \\
\enddata
\end{deluxetable}

\begin{deluxetable}{ccc}
\tablecaption{Crab Nebula fit parameters, 0.5\% systematic errors included\label{tabCrab}}
\tablehead{
\colhead{} & \colhead{2003} &\colhead{2016-18}
}
\startdata
$\alpha_1$ & -2.00 $\pm .02$ & -1.98 $\pm0.02$ \\
$\alpha_2$ & -2.25 $\pm .03$ & -2.32 $\pm0.03$ \\
$E_c (keV)$ & 593 $\pm90$  & 489$\pm55$\\
Norm ($10^{-4} ph/cm^{2}$ s keV)      & 7.45 $\pm 0.24$ & 7.45 $\pm 0.17$\\
100keV Flux($10^{-4}$)& 6.5 & 6.27 \\
Livetime (ks) & 452 & 1386 \\
$\chi_{red}^{2}$ (dof) & 1.9 (40) & 1.36 (40)\\
\tableline
$\alpha_1$ & -1.98 $\pm 0.005$ & -1.98 $\pm0.004$\\
$\alpha_2$ & -2.24 $\pm .03$ & -2.33 $\pm0.03$ \\
$E_c (keV)$ fixed & 500 & 500\\
Norm ($10^{-4} ph/cm^{2}$ s keV)     & 7.68 $\pm 0.035$ & 7.417 $\pm 0.026$\\
100keV Flux($10^{-4}$)& 6.5 & 6.27\\
Livetime (ks) & 452 & 1386 \\
$\chi_{red}^{2}$ (dof) & 1.97 (41) & 1.33 (41)\\
\enddata

\end{deluxetable}

\begin{deluxetable}{cccccccccc}
\tabletypesize{\scriptsize}
\tablecaption{Spectral fit parameters for GS2023+338 during 4 flares.\label{tabV404}}
\tablehead{
\colhead{spectrum}&\colhead{$T_{start}$}&\colhead{$T_{stop}$}&\colhead{$N_H$}&\colhead{Fcomp x$10^{-7}$}&\colhead{kT}&\colhead{$\tau$}&\colhead{NormPl}&\colhead{$\chi^{2}(dof)$}&\colhead{$f_{HE}$} \\
\colhead{Rev. name}&\colhead{MJD-57190}&\colhead{MJD-57190}&\colhead{$10^{22}cm^{-2}$}&\colhead{$ergs/cm^{-2}/s$}&\colhead{keV}&\colhead{}&\colhead{$Ph/cm^2/s/keV$}&\colhead{}&\colhead{\%}
}
\startdata
1554 Rise1&2.189112&2.227478&785$\pm56$ & 0.176 & 23 $\pm1$ & 2.4$\pm0.2$ & 2.7$\pm0.2$  & 138 (153)&65\\
1554 Rise2&2.270975&2.309249&785 & 0.59  & 23  & 1.4$\pm0.1$   & 2.0$\pm0.2$  & 138 (153)&30\\
1554 Max &2.311809&2.350418& 785        &1.5 &23  & 1.5$\pm0.1$ & 6.0$\pm0.4$    & 138 (153)&30\\
1554 Decay1 &2.352642&2.391032& 785        & 0.3 &23  &1.5$\pm0.1$ & 1.2$\pm0.2$   & 138 (153)&32\\
1554 Decay2&2.393012&2.431992& 785        & 0.3   &23          &2.6$\pm0.2$ & 4.1$\pm0.3$   & 138 (153)&63\\
\\
1555 Rise1&4.206403&4.245812& 1300$\pm169$ & 0.375 & 25 & 0.63$\pm0.08$ & 0.5$\pm0.2$ & 219 (150) &15 \\
1555 Rise 2&4.248718&288011& 843$\pm65$ & 1.14 & 25$\pm1$ & 0.78$\pm.06$ & 1.3$\pm0.2$ & 219 (150)& 12\\
1555 Max&4.290397&4.330025& 617$\pm24$ & 4.54 & 25& 0.85$\pm0.06$& 6.9$\pm0.4$& 219 (150)& 15\\
1555 Decay 1 &4.331878&4.371529& 617&1.93 & 25 & 0.77$\pm.06$& 2.8$\pm0.2$& 219 (150) &15\\
1555 Decay 2 &4.373636&4.413335& 0 & 0.16 & 25 & 0.71$\pm.09$ & 0 $\pm0.12$ & 219 (150) & 0\\
\\
1556 Rise 1 &7.028914&7.068914& 1207$\pm50$&0.3  &22.4 $\pm1.5$ & 0.67$\pm.08$ & 0.3$\pm0.1$ & 198 (153)&10\\
1556 Rise 2 &7.070766&7.110742& 1207       &0.51   &22.4          & 0.67         &0.5 $\pm0.1$ & 198 (153)&10\\
1556 Rise 3 &7.112433&7.152455& 1056$\pm100$ &0.52 &   22.4 & 0.48$\pm0.07$ & 0.36$\pm0.1$ &198 (153) & 8\\
1556 Max &7.154748&7.194400&702$\pm34$ & 2.39  & 22.4      & 0.68$\pm0.08$& 2.8 $\pm0.2$ &198 (153) &12\\
1556 Decay 1&7.196229&7.240603& 702  & 0.87 & 22.4 & 0.67 & 1.26$\pm0.13$   &198(153) &15\\
\\
1557 Rise 1 &9.064898&9.071460& 736$\pm24$ &1.2& 27$\pm0.6$ &.78$\pm.04$ & 1.4$\pm0.5$ & 290 (215) &12\\
1557 Rise 2 &9.073417&9.108231& 736 & 0.94 & 27 & 1.15 $\pm.04$ &1.6$\pm0.3$ & 290 (215) &17\\
1557 Max &9.110639&9.144516& 736 & 2.33 &27 & 1.35 $\pm.04$ & 9.0$\pm0.5$ & 290 (215) & 32\\
1557 Decay 1 &9.146380&9.180499& 736 & 1.87 & 27 & 1.72 $\pm.05$ & 10.3$\pm0.5$ & 290 (215) &40 \\
1557 Decay 2&9.218706&9.252432& 736 & 1.96 & 27 & 1.31 $\pm.04$  & 5.9$\pm0.3$ & 290 (215) & 26\\
1557 Decay 3 &9.254910&9.288636& 736  & 1.16 & 27 & 1.14$\pm0.04$ & 2.6$\pm0.2$ & 290 (215) & 21\\
1557 Decay 4 &9.326854&9.360928& 736 & 0.54 &27 & 0.79$\pm.05$ & 0.97 $\pm0.28$ & 290 (215) & 18\\
\enddata
\tabletypesize{\small}
\end{deluxetable}

\begin{deluxetable}{cccc}
\tablecaption{MAXI J1820+070 observation parameters.\label{MAXIobs}}
\tablehead{
\colhead{Revnum} &\colhead{scw interval}&\colhead{$T_{start}$}&\colhead{$T_{stop}$}\\
\colhead{} & \colhead {} & \colhead{MJD-58190} & \colhead{MJD-58190} 
}
\startdata
1931 & 4 - 51 & 3.453081 & 5.567789\\
1932 &10 - 52 & 6.237387 &8.013011\\
1933 & 01 - 51 & 8.765072&10.887870\\
1934 & 10 - 55 & 11.554968 &13.496136\\
1935 & 1 - 18 & 14.039864&14.856437 \\
1936 & 10 - 47 & 16.906067& 18.486923\\
1937 &28 - 41 &20.003440 & 20.578960\\
1938 & 2 - 15 & 22.009841& 22.606113\\
1939 & 2 - 16 & 24.680951& 25.276969\\
1940 &  2 - 15 & 27.328822& 27.924643\\
1941 & 2 - 15 & 29.987514& 30.601865\\
1942 & 18 - 31 & 33.003717& 33.599608\\
1943 & 2 - 15 & 35.305639& 35.919944\\
1944 & 43 - 58 & 38.807687&39.538856 \\
1945 & 2 - 15 & 40.633416& 41.228775\\
1946 & 2 -16 &43.252757& 43.856391\\
1947 & 45 -51 & 47.904203& 47.904203\\
1948 & 28 - 34 & 49.169793& 49.348925\\
1949 & 3 - 15 &51.251982 & 51.555754\\
1950 & 28 - 34 &54.458023 &54.612686 \\
1951 & 2 -15 &56.582536 &56.893681 \\
\enddata
\end{deluxetable}


\begin{thebibliography}{}
\bibitem[Atti\'e et al. (2003)]{att2003}
Atti\'e., D., Cordier, B., Gros, M. et al. \ 2003, \aap, 411, L71
\bibitem[Band et al. (1993)]{band93}
Band, D., Matteson, J., Ford, L. et~al. \ 1993, \apj,  413, 281
\bibitem[Jourdain \& Roques (2009)]{Crab09}
Jourdain, E. \& Roques, J. P. \ 2009,  \apj, 704, 17
\bibitem[Jourdain et al.(2012)]{CygpolarEJ}
Jourdain, E., Roques, J. P., M. Chauvin, M. and Clark, D. J. \ 2012, \apj, 761, 27
\bibitem[Kawamuro et al. (2018)]{Maxi}
 Kawamuro T., Negoro H., Yoneyama T. et al. \ 2018, ATEL 11399, 1
\bibitem[Laurent et al. (2011)]{PL2011}
Laurent, P., Rodriguez, J., Wilms, J et al. \ 2011, Science, 332, 438
\bibitem[Massaro et al. (2000)]{Massaro}
Massaro, E., Cusumano, G., Litterio, M., and Mineo, T. \ 2000, \aap, 361, 895
\bibitem[Roques et al. (2003)]{Roques03}
Roques, J.P., Schanne S., Von Kienlin A. et~al. \ 2003, \aap , 411, L91
\bibitem[Sturner et al. (2003)]{sturner2003} Sturner, S.J., Shrader, C.R.,Weidenspointner, G. et al. 2003, \aap, 411, L81
\bibitem[Trushkin et al. (2018)]{Radio}
Trushkin, S. A., Nizhelskij, N. A., Tsybulev, P. G. and  Erkenov, A. \ 2018, ATEL 11539, 1
\bibitem[Vedrenne et al. (2003)]{Vedrenne03}
Vedrenne, G., Roques, J.P., Schonfelder, V. et~al. \ 2003, \aap, 411, L63
\bibitem[Walton et al.(2017)]{Walton17}
Walton, D., Mooley,K., King A. L. et~al. \ 2017, \apj, 839, 110
\bibitem[Wilson-Hodge et al.(2011)] {WH11}
Wilson-Hodge, C. A., Cherry, M. L.,Case, G.L. et ~al. \ 2011, \apj, 727, L40
\bibitem[Wunderer \& Boggs (2004)]{Wund2004}
Wunderer, T. \& Boggs, S. \ 2004, \url{https://sigma-2.cesr.fr/integral/documents/meetings/4301-MM_SPI_CoIs_271004.pdf}, p177.
\bibitem[Wunderer (2005)]{Wund2005}
Wunderer, T. \ 2005, \url{https://sigma-2.cesr.fr/integral/documents/meetings/4303-CR_SPI-CoIs_March17.pdf}, p71.
\end{thebibliography}
\end{document}